\documentclass[12pt]{article}
\usepackage[utf8]{inputenc}
\usepackage[T1]{fontenc}
\usepackage[margin = 1in]{geometry}
\usepackage{xr-hyper}
\usepackage{ragged2e}
\usepackage{hhline}
\usepackage{varwidth}
\usepackage{graphicx}
\usepackage{pdflscape}
\usepackage{afterpage}
\usepackage{amsmath}
\usepackage{amssymb}
\usepackage{amsfonts}
\usepackage{amsthm}
\usepackage[nodisplayskipstretch, onehalfspacing]{setspace}
\usepackage{subcaption}
\usepackage{adjustbox}
\usepackage{threeparttable}
\usepackage{booktabs}
\usepackage{relsize}
\usepackage{tabularx}
\usepackage{breqn}
\usepackage{soul}
\usepackage[bottom]{footmisc}
\usepackage{longtable} 
\usepackage{hyphenat}
\usepackage{autobreak}
\usepackage{xr}
\theoremstyle{plain}

\newtheorem{proposition}{Proposition}

\newtheorem{assumption}{Assumption}

\newtheorem{remark}{Remark}
\numberwithin{equation}{section}
\numberwithin{assumption}{section}
\numberwithin{theorem}{section}
\numberwithin{algorithm}{section}
\numberwithin{corollary}{section}
\numberwithin{remark}{section}
\numberwithin{proposition}{section}
\numberwithin{lemma}{section}
\numberwithin{table}{section}
\numberwithin{figure}{section}

\allowdisplaybreaks
\usepackage[skip=0pt]{caption}
\usepackage{float}
\floatstyle{plaintop}
\restylefloat{table}

\usepackage{array}
\newcolumntype{H}{>{\setbox0=\hbox\bgroup}c<{\egroup}@{}}

\DeclareGraphicsExtensions{.eps .eps}
\usepackage{epstopdf}

\usepackage[authoryear]{natbib}
\usepackage{xcolor}
\definecolor{marineblue2}{rgb}{0.05,0.1,0.4}
\usepackage{hyperref}
\hypersetup{
unicode=false, 
pdffitwindow=true, 
pdfstartview={FitH}, 
pdftitle={My title}, 
pdfauthor={Author}, 
pdfsubject={Subject}, 
pdfcreator={Creator}, 
pdfproducer={Producer}, 
pdfkeywords={keyword1} {key2} {key3}, 
pdfnewwindow=true, 
colorlinks=true, 
linkcolor=marineblue2, 
citecolor=marineblue2, 
filecolor=marineblue2, 
urlcolor=marineblue2 
}

\makeatletter
\newcommand*{\addFileDependency}[1]{
  \typeout{(#1)}
  \@addtofilelist{#1}
  \IfFileExists{#1}{}{\typeout{No file #1.}}
}
\makeatother

\newcommand*{\myexternaldocument}[1]{%
    \externaldocument{#1}%
    \addFileDependency{#1.tex}%
    \addFileDependency{#1.aux}%
}
\myexternaldocument{web_appendix}
\AtBeginDocument{
\addtolength{\belowdisplayskip}{-1ex}
\addtolength{\belowdisplayshortskip}{-1ex}
\addtolength{\abovedisplayskip}{-1.5ex}
\addtolength{\abovedisplayshortskip}{-1.5ex}
}

\usepackage{appendix} 

\title{The role of parallel trends in event study settings:\\  An application to environmental economics\thanks{First version: January 10, 2020. We thank Brantly Callaway, Jonathan Roth, Julia Schmieder, the Editor, Daniel Millimet, and two anonymous referees for comments and suggestions.}}
\author{Michelle Marcus\\ Vanderbilt University \and Pedro H. C. Sant'Anna\\ Vanderbilt University }

\date{September 3, 2020}

\begin{document}
\maketitle

Difference-in-Differences (DID) research designs usually rely on variation of treatment timing such that, after making an appropriate parallel trends assumption, one can identify, estimate, and make inference about causal effects. In practice, however, different DID procedures rely on different parallel trends assumptions (PTA), and recover different causal parameters. In this paper, we focus on staggered DID (also referred as event-studies) and discuss the role played by the PTA in terms of identification and estimation of causal parameters. We document a ``robustness'' vs. ``efficiency'' trade-off in terms of the strength of the underlying PTA, and argue that practitioners should be explicit about these trade-offs whenever using DID procedures. We propose new DID estimators that reflect these trade-offs and derived their large sample properties. We illustrate the practical relevance of these results by assessing whether the transition from federal to state management of the Clean Water Act affects compliance rates.
\pagebreak

\section{Introduction}
Researchers and policy makers are often interested in evaluating the causal
effect of a given treatment/intervention on an outcome of interest. When
data from randomized control trials related to the causal question of
interest are not available, researchers often rely on \textquotedblleft
natural experiments\textquotedblright\ and make use of
difference-in-differences (DID) methods to estimate the effect of a given
policy. The canonical DID method presumes the existence of two
groups, the treated and the comparison group, two time periods,
pre-treatment and post-treatment periods, such that the comparison group is
not treated in either time period, and the treated group is only treated at
the post-treatment period. Then, one estimates the average treatment effect
among those treated units by comparing the average difference in pre and
post-treatment outcomes of two groups, or, equivalently, by using a two-way
fixed effects regression model with a group and a time fixed effect; see,
e.g., Section 2 of \cite{Lechner2010a} for details about the history of DID procedures.

It is worth stressing that the causal interpretation of the two groups, two
time periods (henceforth $2\times 2$) DID procedure relies on a so-called
parallel trend assumption (PTA): in the absence of the treatment, the
average outcome for the treated and comparison groups would have evolved in
parallel. Such an assumption is well-understood, see e.g. Chapter 5 of \cite{Angrist2009}, Chapter 10 of \cite{Cunningham2018a}, and Section 2 of \cite{Santanna2018}. Importantly, it restricts the average counterfactual
outcome for the treated units \textit{at the post-treatment period }had they
not been subject to the treatment, but it does not directly impose restrictions
on the outcome in pre-treatment periods. In addition, it is worth mentioning
that the PTA is untestable in this $2\times 2$ setup, see, e.g., Chapter 10
of \cite{Cunningham2018a} and Section 4 of \cite{Callaway2018a}.

Although most of the aforementioned points are well-understood in the $%
2\times 2$ setup, in many DID applications, however, there are
more than two time periods, and units can be treated in different points in
time, which leads to multiple treatment groups as well. This
many periods, many groups DID setup is substantially more challenging than
the canonical $2\times 2$ one. For instance, \cite{Abraham2019} (henceforth S\&A), \cite%
{Callaway2018a} (henceforth C\&S), \cite{DeChaisemartin2019} (henceforth dC\&D), and \cite{Goodman-bacon2019}, 
study DID procedures with multiple periods and multiple groups, and each of these
papers rely on \textit{different} types of parallel trends assumptions and/or propose \textit{different} estimators for \textit{%
different} causal parameters of interest. This is in sharp contrast with the 
$2\times 2$ setup, where there is only one type of PTA\footnote{%
In the $2\times 2$ DID setup, the only variation in PTA one observes is
whether it holds unconditionally, or only after conditioning on a vector of
observed characteristics, see, e.g., \cite{Heckman1998}, \cite{Abadie2005},
and \cite{Santanna2018}. This is not the type of variation of the PTA we are
referring to.} and the main parameter of interest is the average treatment
effect among the treated units. It is also worth stressing that in this more complex DID setup treatment effect estimates based on
two-way fixed effects (TWFE) regression models can only be interpreted as
weighted averages of causal effects, and, perhaps even more problematic, that
some of these weights can be negative, see, e.g., S\&A, \cite{Borusyak2017}, dC\&D, and \cite{Goodman-bacon2019}; see also \cite{Laporte2005}, \cite{Wooldridge2005}, \cite{Chernozhukov2013}, and  \cite{Gibbons2018} for earlier related results based on (one-way) fixed-effect estimators. As so, policy evaluations based on TWFE linear regression models can be misleading, especially when treatment effects are dynamic.

In this paper, we attempt to clarify the role played by different parallel
trends assumptions in DID setups with many time periods and many treatment
groups. Towards this goal, we revisit S\&A, C\&S, and dC\&D,
and discuss their main differences in terms of parallel trends assumptions and parameters of interest --- we focus on these three papers because their primary concern is how one can identify causal parameters of interest without imposing restrictions beyond a PTA\footnote{See also \cite{Athey2018}, \cite{Goodman-bacon2019}, \cite{Arkhangelsky2018}, \cite{Borusyak2017}, \cite{Ferman2018}, and \cite{Rambachan2019} for other recent contributions to the DID literature.}. We exclusively focus on DID settings with staggered adoption designs and binary treatments. By doing so, we can compare the PTA and parameters of interest discussed in S\&A, C\&S, and dC\&D in a more direct manner.

We show that the PTA invoked by S\&A and dC\&D (i) not only restricts counterfactual trends after the treatment, but also imposes parallel pre-treatment trends, and (ii) imposes that every \emph{individual} group that is not-yet treated by time $t$ can be used as a valid comparison group for those earlier-treated units, at time $t$.  C\&S, on the other hand, considers two different PTAs, one that relies on using ``never-treated'' units as a comparison group, and one that uses not-yet-treated units as valid comparison groups for the earlier-treated units. Interestingly, both PTAs considered by C\&S are, at least technically speaking, weaker than the PTA invoked by S\&A and dC\&D, as they either do not restrict pre-treatment trends, or, when they do, these restrictions are potentially less demanding. Although these PTAs differ in their ``strength'', we show that they all can be used to recover the same variety of average treatment effects measures.

Overall, we argue that, in practice, one should be explicit about the type of PTA invoked in the DID analysis. On top of adding transparency and objectivity to the analysis, see, e.g., \cite{Rubin2007, Rubin2008}, we stress that the choice of the parallel trends assumption can also help in selecting the most appropriate estimator for a given parameter of interest. For instance, in situations where one is comfortable with a ``stronger'' PTA, we show that one can exploit the fact that such PTAs lead to \emph{overidentification}, and then use the generalized method of moments (GMM) framework to form more efficient treatment effect estimators than those currently available in the literature; see Proposition \ref{Thm:1}. 
Another consequence of adopting the GMM framework is that it is relatively straightforward to test for the credibility of a ``stronger'' parallel trends assumption by conducting a classical Hansen-Sargan J-test. To the best of our knowledge, this paper is the first to make this simple, but important observation.

In many other situations, however, we expect that researchers will not be a priori comfortable with a ``stronger'' version of the PTA, as it may impose more restrictions on the data than those strictly required for identification of treatment effect parameters. Indeed, when the number of groups and time periods is moderate, the number of restrictions implied by the ``stronger'' PTA can be close to the number of observations available in the data. In such cases, it may be reasonable to favor ``weaker'' versions of the PTA. When a sufficiently large ``never-treated'' group is available, researchers can use the easy-to-implement nonparametric DID estimators based on sample means proposed by C\&S. When an appropriate ``never-treated'' group is unavailable, we show that one can rely on an alternative ``weaker'' PTA and use a simple plug-in DID estimator that differs from the ones considered by S\&A, C\&S, and dC\&D. We show that this new DID estimator is consistent and asymptotically normal, and also describe a bootstrap procedure to conduct inference that is robust against multiple-testing problems. Interestingly, this new DID estimator does not rely on restricting pre-treatment trends, and, at the same time, exploits data from all available groups in the given application. On the other hand, both this newly proposed DID estimator and the one proposed by C\&S are, in general, less efficient than the GMM estimator, which relies on more stringent assumptions. To the best of our knowledge, we are the first to document this ``robustness'' versus ``efficiency'' trade-off in terms of the strength of the underlying PTA invoked in DID setups.

We illustrate the practical relevance of the aforementioned observations by
revisiting \cite{grooms2015}. We examine the effect of the transition from federal to state management of the Clean Water Act (CWA) on violation rates. Similarly to \cite{grooms2015}, we find that the transition from federal to state control has little to no effect on violation rates --- this result is robust across different parallel trends assumptions and different causal parameters of interest. 

Next, like \cite{grooms2015}, we also analyze whether states with a long prevalence of corruption see a large decrease in the violation rate after authorization relative to states without corruption. \cite{grooms2015} uses a dynamic TWFE (event-study) linear regression model, and finds strong evidence that violation rates decreased more in more corrupt states than in less corrupt states after the transition to state control. However, given that \cite{grooms2015} focuses exclusively on TWFE-type estimators, it is not clear what kind of PTA is actually being made in the analysis. Here, we show how it can be beneficial to separate the analysis into two steps: (i) identification and the relevance of the PTA, and (ii) data analysis and estimation procedures. By proceeding in this manner, we find that the conclusion that violation rates dropped more in more corrupt states than less corrupt states depends on the type of PTA imposed. For instance, when one assumes that, in the absence of treatment, the counterfactual outcome trends differ depending on whether a state has a long prevalence of corruption or not (``corruption-specific trends''), we find essentially no evidence that the treatment effects vary depending on whether a state is more or less corrupt. On the other hand, if one assumes an alternative PTA such that one can use averages of both corrupt and non-corrupt states as valid comparison groups, we find evidence that more corrupt states see a larger decrease in the violation rate after authorization than less corrupt states, just like the original findings of \cite{grooms2015}. As ``corruption'' is not randomly assigned, we believe that allowing for corruption-specific trends is the most natural identification set up in this context. These conflicting findings highlight the importance of explicitly stating the underlying PTA invoked in the exercise.

The rest of this paper is organized as follows. In Section \ref{sec2}, we present the general framework, compare the different PTAs using a stylized example, and describe the different parameters of interest considered by S\&A, C\&S, and dC\&D. In Section \ref{sec3}, we discuss the testability of the PTAs and the practical considerations a researcher might take into account when choosing a PTA and a DID estimator. Section \ref{sec:GMM} describes how one can use generalized method of moments (GMM) framework to form more efficient treatment effect estimators when the chosen PTA leads to overidentification. Section \ref{sec:Alternative} presents a new easy-to-compute DID estimator based on an alternative ``weaker'' PTA than what has been previously seen in the literature. Finally, Section \ref{sec:Application} presents the empirical application, and we conclude in Section \ref{sec:Conclusion}. Proofs and additional results are available at the Web Appendix, at \href{https://pedrohcgs.github.io/files/Marcus_SantAnna_2020_webAppendix.pdf}{https://pedrohcgs.github.io/files/Marcus_SantAnna_2020_webAppendix.pdf}.

\section{Difference-in-differences with multiple time periods\label{sec2}}

\subsection{Framework \label{sec:framework}}

We first introduce the notation we use throughout the paper, which
resembles that adopted by C\&S. We consider the case with 
$\mathcal{T}$ periods and denote a particular time period by $t$ where $%
t=1,\ldots ,\mathcal{T}$. In the canonical DID setup, $\mathcal{T}=2$ and no
one is treated in period 1. Let $D_{t}$ be a binary variable equal to one if
a unit is treated in period $t$ and equal to zero otherwise. Also,
define $G_{g}$ to be a dummy variable that is equal to one if a unit
is first treated in period $g$, and define $C$ as a dummy variable that is
equal to one for units who are not treated in any period $\mathcal{T}$.
For each unit, exactly one of the $G_{g}$ or $C$ is equal to one.
Finally, let $Y_{t}\left( 1\right) $ and $Y_{t}\left( 0\right) $ be the
potential outcomes at time $t$ with and without treatment, respectively. The
observed outcome in each period can be expressed as $Y_{t}=D_{t}Y_{t}\left(
1\right) +\left( 1-D_{t}\right) Y_{t}\left( 0\right) .$ 

Henceforth, we refer to \textquotedblleft groups\textquotedblright\ as the
group associated with the time a unit is first treated. Throughout the paper, we maintain the following
assumptions.

\begin{assumption}[Sampling]
\label{ass:iid} $\{Y_{i1},Y_{i2},\ldots Y_{i\mathcal{T}},D_{i1},D_{i2},%
\ldots ,D_{i\mathcal{T}}\}_{i=1}^{n}$ is independent and identically
distributed $(iid)$.
\end{assumption}

\begin{assumption}[Staggered treatment design]
\label{ass:stag} For $t=2,\ldots ,\mathcal{T}$, 
\begin{equation*}
D_{t-1}=1\text{ implies that }D_{t}=1
\end{equation*}
\end{assumption}

\begin{assumption}[No Anticipation]
\label{ass:anticip} For all $t=1,\ldots ,\mathcal{T}$, $g=2,\ldots ,\mathcal{T}$ such that $t<g$,
\begin{equation*}\mathbb{E}\left[ \left. Y_{it}
\right\vert G_{g}=1\right] = \mathbb{E}\left[ \left. Y_{it}\left( 0\right)
\right\vert G_{g}=1\right]
\end{equation*}

\end{assumption}

\begin{assumption}[Overlap]
\label{ass:overlap} $P\left( G_{1}=1\right) =0$ and, for some $\epsilon>0$, and all $g=2,\ldots ,%
\mathcal{T}$, $P\left( G_{g}=1\right) >\epsilon$.
\end{assumption}

Assumption \ref{ass:iid} implies that we are considering the case of panel data. The discussions related to the case where only repeated cross-section
data are available follows similar arguments and is omitted to avoid
repetition. Assumption \ref{ass:iid} does not restrict the temporal dependence across outcomes, though it relies on ``large n, fixed t'' panel data. Assumption \ref{ass:iid} also rules out covariates; we only
impose this simplification to allow for a more direct comparison between the
proposals of S\&A, C\&S, and dC\&D; we refer the reader to C\&S for a detailed discussion about flexibly accommodating covariates into the DID analysis.

Assumption \ref{ass:stag} imposes that treatment is \textquotedblleft
irreversible\textquotedblright , i.e., once a unit is treated at time $t-1$,
it is \textquotedblleft forever\textquotedblright\ treated. This assumption
is usually referred to as staggered treatment adoption in the DID literature.
We interpret this assumption as if units that experience treatment are
forever affected by this experience, and do not \textquotedblleft
forget\textquotedblright\ about it.\footnote{When treatment can
\textquotedblleft turn on\textquotedblright\ and later \textquotedblleft
turn off\textquotedblright , one usually needs to augment the potential outcome notation 
to analyze the effect of a given treatment path, see, e.g. \cite%
{Han2019}. When one is interested 
only in an average of the instantaneous effects of the policy among all units that switch treatments, one can bypass some
of this complications by imposing a ``no carryover assumption'', see, e.g., dC\&H.} We emphasize that, by
imposing Assumption \ref{ass:stag}, we are able to directly compare the DID
contributions of S\&A, C\&S, and dC\&D.

Assumption \ref{ass:anticip} implies that there is no anticipatory response to treatment for those units that are eventually treated. This assumption is standard in the DID literature, though many times it is only implicitly imposed. Such an assumption can be violated if units foresee that they will be treated and change their behavior before the treatment takes place.\footnote{When the researcher is worried about anticipatory effects, this can be circumvented by simply redefining $g$ to denote the period in which anticipatory effects begin. However, this may require strengthening other assumptions; see C\&S for a discussion.}

Assumption \ref{ass:overlap} imposes that no unit is treated in the first
time period, and that a new set of units are treated in each time period with
a strictly positive probability. If there is an ``always treated'' group, i.e., units that are already treated in the first time period, we drop those observations because neither the data nor parallel trends assumptions for $Y_t(0)$ provide information information to identify the average treatment effect for this group.\footnote{In this paper, we attempt to only make parallel trends assumption about the evolution of $Y_t(0)$, and remain agnostic about the trends for $Y_t(1)$. When one is willing to impose parallel trends for $Y_t(1)$, too, then we can leverage the existence of an ``always treated'' group to form alternative parameters of interest, though such an assumption further restricts treatment effect heterogeneity. We leave a detailed discussion about this case for future research.} We assume that new sets of units are treated in each time period only for notation convenience. Also, note that Assumption \ref{ass:overlap} accommodates, but does not require, that there is a \textquotedblleft never treated\textquotedblright\ group
available. 

Next, we revisit S\&A, C\&S, and dC\&D, paying particular
attention to their PTA, underlying parameter of interest, and how one can estimate these parameters. When
presenting these results, we impose Assumptions \ref{ass:iid} - \ref%
{ass:overlap}, which may result in slight changes of notation when compared to
their original statements. In terms of notation, we follow C\&S and,
whenever possible, attempt to express the different parameters of interest
in terms of functionals of ``group-time average treatment effects'', i.e., the average treatment effect at time $t$, for
those units first treated at time $g$,\footnote{S\&A refers to $ATT\left( g,t\right)$ as cohort-specific
average treatment effects on the treated, though they express it in terms of event-time $t-g$, i.e., the time elapsed since
treatment started.} 
\begin{eqnarray}
ATT\left( g,t\right) &\equiv& \mathbb{E}\left[ \left. Y_{it}\left( 1\right) \right\vert G_{g}=1\right] - \mathbb{E}\left[ \left. Y_{it}\left( 0\right) \right\vert G_{g}=1\right].  \label{ATT_gt} \\
&=& \alpha _{g,t}\left(1\right) - \alpha _{g,t}\left( 0\right) \nonumber
\end{eqnarray}

To convey the discussion in an easy-to-understand manner, we consider a simple, stylized example. Assume that we observe $Y_{it}$ for a sample of units $i=1,\dots ,n$ in four
time periods, $t=1,2,3,4$. Some units are first treated at time $3$ ($%
G_{i3}=1)$, others at time $4$, ($G_{i4}=1$), and the remaining units are not treated
in the entire observation window $(C_{i}=1)$. Once a unit $i$ is treated at
time $g$, it remains treated for all time periods $t\geq g$. Let $W=\left( Y_{1},Y_{2},Y_{3},Y_{4}, G_{3},G_{4},C\right) ^{\prime }$, and assume that we observe a random
sample $\left\{ W_{i}\right\} _{i=1}^{n}$ of $W$.



\subsection{The different parallel trends assumptions \label{sec:pta}}
In this subsection, we present the three different parallel trends assumptions considered by S\&A, C\&S, and dC\&D. We start by describing each PTA conceptually and then make use of the stylized example to highlight the key differences between these assumptions.

We first present the PTA assumption invoked by S\&A and dC\&D (see, Assumption 1 in S\&A and Assumption 5 in dC\&D).

\begin{assumption}[Parallel trends assumption across all time periods and all
groups]
\label{ass:dCdH-PTA} For all $t=2,\ldots ,\mathcal{T}$, all $g=2,\ldots ,%
\mathcal{T}$, 
\begin{equation*}
\mathbb{E}\left[ \left. Y_{t}\left( 0\right) -Y_{t-1}\left( 0\right)
\right\vert G_{g}=1\right] =\mathbb{E}\left[ \left. Y_{t}\left( 0\right)
-Y_{t-1}\left( 0\right) \right\vert C=1\right] =\mathbb{E}\left[ Y_{t}\left(
0\right) -Y_{t-1}\left( 0\right) \right] ,
\end{equation*}%
where the first equality holds only when there exists a \textquotedblleft
never-treated\textquotedblright\ group.
\end{assumption}

Assumption \ref{ass:dCdH-PTA} states that, in the absence of treatment, the expectation of the outcome of interest follows the same path in all groups and in all time periods available in the data. Although fairly
intuitive, such an assumption imposes important restrictions on the data (when combined with Assumptions \ref{ass:iid} - \ref{ass:overlap}). In particular, Assumption \ref{ass:dCdH-PTA} imposes a parallel pre-trends condition across all treatment groups, and, as a consequence, allows one to use any \emph{individual} group that has not yet been treated by time $t$ (units with $G_s=1$, $s>t$) as a valid comparison group for those units already treated by time $t$.

To visualize these restrictions, let us consider our stylized example. Under Assumptions \ref{ass:iid} - \ref{ass:overlap}, the PTA \ref{ass:dCdH-PTA} can be written as the following seven moment conditions:%
\begin{eqnarray}
\alpha _{3,3}\left( 0\right)  &=&\mathbb{E}\left[ Y_{3}-Y_{2}|C=1\right] +%
\mathbb{E}\left[ Y_{2}|G_{3}=1\right] ,  \label{mom:3_3_1} \\
\alpha _{3,3}\left( 0\right)  &=&\mathbb{E}\left[ Y_{3}-Y_{2}|G_{4}=1\right]
+\mathbb{E}\left[ Y_{2}|G_{3}=1\right] ,  \label{mom:3_3_2} \\
\alpha _{3,4}\left( 0\right)  &=&\mathbb{E}\left[ Y_{4}-Y_{3}|C=1\right]
+\alpha _{3,3}\left( 0\right) ,  \label{mom:3_4_1} \\
\alpha _{4,4}\left( 0\right)  &=&\mathbb{E}\left[ Y_{4}-Y_{3}|C=1\right] +%
\mathbb{E}\left[ Y_{3}|G_{4}=1\right] ,  \label{mom:4_4_1} \\
\mathbb{E}\left[ Y_{2}-Y_{1}|G_{3}=1\right]  &=&\mathbb{E}\left[
Y_{2}-Y_{1}|C=1\right] ,  \label{mom:trends1} \\
\mathbb{E}\left[ Y_{2}-Y_{1}|G_{3}=1\right]  &=&\mathbb{E}\left[
Y_{2}-Y_{1}|G_{4}=1\right]   \label{mom:trends2} \\
\mathbb{E}\left[ Y_{3}-Y_{2}|G_{4}=1\right]  &=&\mathbb{E}\left[
Y_{3}-Y_{2}|C=1\right] .  \label{mom:trends3}
\end{eqnarray}%
These moment restrictions highlight all restrictions PTA \ref{ass:dCdH-PTA} impose on the data. First, the moment conditions (\ref{mom:3_3_1}) and (\ref{mom:3_3_2}) formalize the
notion that the evolution of the outcome for the \textquotedblleft
never-treated\textquotedblright\ and \textquotedblleft
late-treated\textquotedblright\ units can be used to identify $\alpha_{3,3}(0) $, which in
turn, would allow one to identify $ATT\left( 3,3\right). $\footnote{%
Recall that, for $g\geq t,$ $ATT\left( g,t\right)  = \alpha _{g,t}\left( 1\right) - \alpha _{g,t}\left( 0\right) = \mathbb{E}\left[
Y_{t}|G_{g}=1\right] -\mathbb{E}\left[ Y_{t}\left( 0\right) |G_{g}=1\right].
$ Thus, given that $\mathbb{E}\left[ Y_{t}|G_{g}=1\right] $ is estimable
from the data, one only needs to identify $\alpha _{g,t}\left( 0\right)$ in order to recover $ATT\left( g,t\right) $ from
the data.} An empirically important implication of this observation is that 
\emph{any linear combination }of\textit{\ }$\mathbb{E}\left[ Y_{3}-Y_{2}|C=1%
\right] $ and $\mathbb{E}\left[ Y_{3}-Y_{2}|G_{4}=1\right] $ can be used to
impute $\alpha _{3,3}\left( 0\right) $. Given that the \textquotedblleft
never-treated\textquotedblright\ units are the only units that have not yet
experienced treatment at time $t=4$, they form the only group that can be
used to recover $\alpha _{3,4}\left( 0\right) $ and $\alpha _{4,4}\left(
0\right) $--- this notion is formalized by the moment restrictions (\ref%
{mom:3_4_1}) and (\ref{mom:4_4_1}). Finally, (\ref{mom:trends1})-(\ref%
{mom:trends3}) impose a parallel \textquotedblleft pre-trends\textquotedblright\
condition, i.e., that the evolution of the outcome before treatment occurs is
the same across all groups. Note that the moment condition (\ref{mom:trends3}%
) is a linear combination of the moment conditions (\ref{mom:3_3_1}) and (%
\ref{mom:3_3_2}), so (\ref{mom:trends3}) is redundant in the aforementioned system of equations. Nonetheless, this observation allows one to conclude that, by assuming that both never-treated units and the units that are not-yet-treated at time $t=3$ can be used as valid comparison groups for the units first treated at time $t=3$, one is imposing the parallel pre-trends condition (\ref{mom:trends3}). Of course, the reverse argument is also true, highlighting that parallel pre-trends across groups may have important identification content; we further discuss this in Section~\ref{sec3}.

Given that we now have a better understanding of the PTA invoked by S\&A and dC\&D, we turn our attention to the PTA invoked by C\&S. In fact, C\&S consider two different PTA depending on whether a ``never treated'' group is available or not (see Assumptions 4 and 5 in C\&S).

\begin{assumption}[Parallel trends assumption based on \textquotedblleft
never treated\textquotedblright\ units]
\label{ass:CS-PTA1} For all $g,t=2,\ldots ,\mathcal{T}$, $g=2,\ldots ,\mathcal{%
T}$, such that $t\geq g$, 
\begin{equation*}
\mathbb{E}\left[ \left. Y_{t}\left( 0\right) -Y_{t-1}\left( 0\right)
\right\vert G_{g}=1\right] =\mathbb{E}\left[ \left. Y_{t}\left( 0\right)
-Y_{t-1}\left( 0\right) \right\vert C=1\right] .
\end{equation*}
\end{assumption}

\begin{assumption}[Parallel trends assumption based on \textquotedblleft
not-yet treated\textquotedblright\ units]
\label{ass:CS-PTA2} For all $g,s,t =2,\ldots ,\mathcal{T}$, such that $t\geq g$, $s\geq t$, 
\begin{equation*}
\mathbb{E}\left[ \left. Y_{t}\left( 0\right) -Y_{t-1}\left( 0\right)
\right\vert G_{g}=1\right] =\mathbb{E}\left[ \left. Y_{t}\left( 0\right)
-Y_{t-1}\left( 0\right) \right\vert D_{s}=0\right] .
\end{equation*}
\end{assumption}

The difference between the parallel trends assumptions \ref{ass:CS-PTA1} and %
\ref{ass:CS-PTA2} is that the former uses the \textquotedblleft never
treated\textquotedblright\ units as a fixed comparison group, whereas the
latter allows one to use averages of different groups of units that are not-yet treated by time $t$ as
a comparison group. At first sight, it may not be clear whether the PTAs \ref{ass:CS-PTA1} and \ref{ass:CS-PTA2} also restrict pre-trends as the PTA \ref{ass:dCdH-PTA} does. 

In order to compare these PTAs, it is illustrative to focus our attention to the stylized example where we again pre-impose Assumptions \ref{ass:iid} - \ref{ass:overlap}.
In this context, the PTA \ref{ass:CS-PTA1} imposes the following three moment restrictions:%
\begin{eqnarray}
\alpha _{3,3}\left( 0\right)  &=&\mathbb{E}\left[ Y_{3}-Y_{2}|C=1\right] +%
\mathbb{E}\left[ Y_{2}|G_{3}=1\right] ,  \label{mom:33} \\
\alpha _{3,4}\left( 0\right)  &=&\mathbb{E}\left[ Y_{4}-Y_{3}|C=1\right]
+\alpha _{3,3}\left( 0\right) ,  \label{mom:34} \\
\alpha _{4,4}\left( 0\right)  &=&\mathbb{E}\left[ Y_{4}-Y_{3}|C=1\right] +%
\mathbb{E}\left[ Y_{3}|G_{4}=1\right] .  \label{mom:44}
\end{eqnarray}%
As is evident from (\ref{mom:33})-(\ref{mom:44}), the PTA \ref{ass:CS-PTA1} does not restrict pre-trends across groups, and does not presume that ``later treated'' units can be used as a valid comparison group for ``early treated'' units. Although the moment conditions (\ref{mom:33}), (\ref%
{mom:34}), and (\ref{mom:44}) are respectively the same as (\ref{mom:3_3_1}), (\ref%
{mom:3_4_1}), and (\ref{mom:4_4_1}), it does not impose the moment restrictions (\ref{mom:3_3_2}), (\ref{mom:trends1}), and (\ref{mom:trends2}) imposed by the PTA \ref{ass:dCdH-PTA}. Therefore, one can reasonably argue that the PTA \ref{ass:CS-PTA1} is \textquotedblleft weaker\textquotedblright\ then the PTA \ref{ass:dCdH-PTA}.

Next, we describe the PTA \ref{ass:CS-PTA2}, which, in the context of our stylized example, imposes the following moment restrictions:
\begin{eqnarray}
\alpha _{3,3}\left( 0\right)  &=&\mathbb{E}\left[ Y_{3}-Y_{2}|D_{3}=0\right]
+\mathbb{E}\left[ Y_{2}|G_{3}=1\right] ,  \label{mom:33_ny} \\
\alpha _{3,3}\left( 0\right)  &=&\mathbb{E}\left[ Y_{3}-Y_{2}|D_{4}=0\right]
+\mathbb{E}\left[ Y_{2}|G_{3}=1\right] ,  \label{mom:33_new_ny} \\
\alpha _{3,4}\left( 0\right)  &=&\mathbb{E}\left[ Y_{4}-Y_{3}|D_{4}=0\right]
+\alpha _{3,3}\left( 0\right) ,  \label{mom:34_ny} \\
\alpha _{4,4}\left( 0\right)  &=&\mathbb{E}\left[ Y_{4}-Y_{3}|D_{4}=0\right]
+\mathbb{E}\left[ Y_{3}|G_{4}=1\right] .  \label{mom:44_ny}
\end{eqnarray}%
where $D_{it}=1$ if a unit $i$ is treated by time $t$, and equal to zero
otherwise. In the context of our stylized example, $D_{4}=0$ if and only if $C=1
$, implying that (\ref{mom:33_new_ny})-(\ref{mom:44_ny}) are equivalent to (\ref{mom:33})-(\ref{mom:44}), respectively. Thus, from this simple observation, we can conclude that the PTA \ref{ass:CS-PTA2} is \textquotedblleft stronger\textquotedblright\ then PTA \ref{ass:CS-PTA1}, as the latter does not involve the moment restriction (\ref{mom:33_ny}). 

To compare the PTA \ref{ass:CS-PTA2} with the PTA \ref{ass:dCdH-PTA}, we need to understand the implications of adding the moment restriction (\ref{mom:33_ny}) to the other moment restrictions implied by PTA \ref{ass:CS-PTA1}. Note that when we combine (\ref{mom:33_ny}) with (\ref{mom:33_new_ny}), we have that $\mathbb{E}\left[ Y_{3}-Y_{2}|D_3=0\right] = \mathbb{E}\left[ Y_{3}-Y_{2}|D_4=0\right]$, which in our example is the same as $\mathbb{E}\left[ Y_{3}-Y_{2}|D_3=0\right] = \mathbb{E}\left[ Y_{3}-Y_{2}|C=1\right]$. Given that $D_{3}=0$ if and only if either $G_{4}=1$ or $C=1$, it follows that 
\begin{eqnarray*}
\mathbb{E}\left[ Y_{3}-Y_{2}|D_3=0\right] &=& \mathbb{E}\left[ Y_{3}-Y_{2}|C=1\right]\\
&\iff& \\
\mathbb{E}\left[ Y_{3}-Y_{2}|G_{4}=1\right] &=& \mathbb{E}\left[ Y_{3}-Y_{2}|C=1\right].
\end{eqnarray*}
Thus, by exploiting this simple but subtle observation, we can conclude that the moment restrictions implied by the PTA \ref{ass:CS-PTA2}, (\ref{mom:33_ny})-(\ref{mom:44_ny}), are \emph{equivalent} (\ref{mom:3_3_1})-(\ref{mom:4_4_1}), a subset of the moment restrictions implied by the PTA \ref{ass:dCdH-PTA}. Importantly, and in contrast with the PTA \ref{ass:CS-PTA1}, the PTA \ref{ass:CS-PTA2} does rule out non-parallel pre-trends for \emph{some} groups and pre-treatment periods,\footnote{The PTA \ref{ass:CS-PTA2} does not restrict pre-trends involving time periods before the first unit is treated, and does not restrict pre-trends for the earliest treatment group.} though, technically, it is still weaker than the PTA \ref{ass:dCdH-PTA} as the latter completely rules out any type of non-parallel pre-trends.

In summary, from the discussion presented above, one can conclude that the PTA \ref{ass:CS-PTA1} does not restrict pre-trends and is weaker than the PTA \ref{ass:dCdH-PTA} and \ref{ass:CS-PTA2}, though it requires the existence of a ``never treated'' group. In addition, the PTA \ref{ass:CS-PTA2} is arguably weaker then the PTA \ref{ass:dCdH-PTA}, as the latter restricts all pre-trends in all pre-treatment periods, while the former does not restrict pre-trends involving time periods before the first unit is treated. This can be practically relevant in applications where data are available on many time periods before the first group of units is treated.

\subsection{Parameters of interest}
In this subsection, we discuss the different parameters of interest that may arise when one deviates from the canonical $2\times 2$ DID setting. Before presenting the parameters of interest considered by S\&A, C\&S, and dC\&D, it is worth stressing the potential pitfalls associated with the commonly used TWFE regression specifications.

\subsubsection{Pitfalls of TWFE regression specifications\label{sec:pit}}
As \cite{Borusyak2017}, dC\&D and \cite{Goodman-bacon2019} point out, one of the most popular specifications in this many periods and many groups DID setting is the following TWFE regression specification,\footnote{All the results remain the same if one replaces $\alpha_{g}$ with $\alpha_{i}$, an unit-specific fixed effect. We prefer to include $\alpha_{g}$ as it closely resemble the canonical DID regression specification.}%
\begin{equation}
Y_{it}=\alpha _{g}+\alpha _{t}+\beta _{fe}D_{it}+u_{it},  \label{eq:TWFE-1}
\end{equation}%
where $\alpha _{g}$ and $\alpha _{t}$ are group and time fixed effects,
respectively, and $u_{it}$ is an idiosyncratic error term. Although practitioners often consider $\beta_{fe}$ to be a main parameter of interest, these aforementioned papers show that, when treatment effects are allowed to be heterogeneous across groups and time periods, $\beta_{fe}$ can only be interpreted as a weighted average of treatment effects, and, perhaps even more problematic, some of these weights can be negative; see also \cite{Laporte2005}, \cite{Wooldridge2005}, \cite{Chernozhukov2013}, and  \cite{Gibbons2018} for earlier related results based on (one-way) fixed-effect estimators. As so, interpreting estimates of $\beta_{fe}$ as sensible causal summary parameters can lead to misleading conclusions about the policy effectiveness.

Moreover, the negative (and non-intuitive) weighting problem is not specific to (\ref{eq:TWFE-1}). dC\&D show that it also applies to the first-difference specification. In addition, S\&A show the non-convex weighting problem extends to many variations of the dynamic TWFE regression specification\begin{equation}
Y_{it}=\alpha_{i}+\alpha_{t}+\sum_{e =-K}^{-1}\beta _{e }1\left\{
t-G_{i}+1=e \right\} +\sum_{e =1}^{L}\beta _{e }1\left\{ t-G_{i}+1=e
\right\} +v_{it},  \label{eq:TWFE2}
\end{equation}%
where $G_{i}$ is the time a unit $i$ is first treated (equal to infinity if unit $i$ is ``never-treated''), $t-G_{i}+1$ is the
``event time'' , i.e., the number of time
periods a unit has been treated, $1\left\{ t-G_{i}+1=e \right\} $ is an indicator
for unit $i$ being treated for $e$ time periods. Taken together, these results suggest that the common practice of attaching sensible causal interpretation to the coefficients of TWFE regression models is not, in general, warranted.

\subsubsection{More sensible treatment effect parameters and their estimators\label{sec:parameters}}

Given the potential pitfalls associated with traditional estimation procedures, S\&A, C\&S and dC\&D propose different estimators for different treatment effect parameters. In this subsection, we review these procedures and highlight their differences.

Given that traditional estimation procedures do not recover easy to interpret causal parameters without further restricting treatment effect heterogeneity, S\&A, C\&S and dC\&D propose different estimators for different treatment effect parameters. In this subsection, we review these procedures.

dC\&D focuses on an instantaneous treatment effect measure across all ``ever treated'' groups. More precisely, dC\&D is mainly interested in estimating 
\begin{equation}
\delta ^{S}\equiv \mathbb{E}\left[ \frac{\sum_{i=1}^{n}\sum_{t=2}^{\mathcal{T%
}}G_{ig}\cdot \left( Y_{it}\left( 1\right) -Y_{it}\left( 0\right) \right) }{%
\sum_{i=1}^{n}\sum_{t=2}^{\mathcal{T}}G_{it}}\right] ,  \label{eq:ATS}
\end{equation}%
the average of the treatment effect at the time when a group starts receiving the treatment, across all groups that become treated at some point (see Section 4 of dC\&D).\footnote{This is the case with staggered treatment adoption. In more general treatment adoption setups, the parameter of interest considered by dC\&H differs from $\delta^{S}$ as defined above.}

dC\&D also proposes an easy-to-implement estimator for $\delta ^{S}$. To better understand their estimator, let 
\begin{equation}
\widehat{ATT}_{ny}\left( g,t\right) =\frac{n^{-1}\sum_{i=1}^{n}G_{ig}\left(
Y_{it}-Y_{ig-1}\right) }{n^{-1}\sum_{i=1}^{n}G_{it}}-\frac{%
n^{-1}\sum_{i=1}^{n}\left( 1-D_{it}\right)\left( 1-G_{ig}\right) \left( Y_{it}-Y_{ig-1}\right) }{%
n^{-1}\sum_{i=1}^{n}\left( 1-D_{it}\right)\left( 1-G_{ig}\right) }
\label{eq:att_ny}
\end{equation}%
be a DID estimator for (\ref{ATT_gt}) that uses not-yet treated units by time $t$ as a
comparison group for treatment group $g$, at time $t$. Consider the estimator for the probability of a unit being in group $g$ given that
it is among the units that are treated for at least $e=t-g+1$ periods given by
\begin{equation}
\widehat{w}(g;e) \equiv \widehat{P}\left( G_{g}=1| \text{Treated for $\ge$ e periods}\right) =\frac{N_{g\cap \geq e}}{%
N_{\geq e}},
\label{eq:dyn-weights}
\end{equation}%
where $N_{g\cap \geq e}$ denotes the number of observations in group $g$ among
those units that have been treated for at least $e$ periods, and $N_{\geq e}$ is
the number of units who have been treated for at least $e$ periods. dC\&D then show that, under the PTA \ref%
{ass:dCdH-PTA} and some additional regularity conditions,  
\begin{equation}
\widehat{\delta }^{S}=\sum_{g=2}^{\mathcal{T}}\widehat{P}\left( G_{g}=1|\text{Treated for $\ge$ 1 period}\right) \cdot \widehat{ATT}_{ny}\left( g,g\right),
\label{eq:ATS_est}
\end{equation}%
is an unbiased estimator of $\delta ^{S}$, and, as (effective) sample size
grows, $\widehat{\delta }^{S}$ is also consistent and asymptotically normal.

From the discussion above, it is evident that (\ref{eq:ATS_est}) is a well-defined estimator for the easy-to-interpret causal parameter of interest $\delta ^{S}$ as defined in (\ref{eq:ATS}). On the other hand, $\widehat{\delta }^{S}$ is, by design, only suitable to summarize instantaneous treatment effects. Hence, when one is interested in treatment effect dynamics, one needs to consider alternative causal parameters of interest.

A particular way of considering more general parameters of interest that are able to capture richer sources of treatment effect heterogeneity is to follow C\&S, and center the analysis on the average treatment effect at time $t$, for those units first treated at
time $g$, $ATT\left( g,t\right)$ as defined in (\ref{ATT_gt}). By doing so, one can highlight different sources of treatment effect heterogeneity. For instance, one can look at how the $ATT\left( g,t\right)$ for particular group $g$ evolves over time, which would allow one to study group-specific treatment effect dynamics. Alternatively, one can form different weighted averages of the  $ATT\left( g,t\right)$ that are able to summarize overall treatment effects. Examples of these weighted averages include $(i)$ a ``simple'' average of the $ATT\left( g,t\right)$,
\begin{equation}
ATT^{simple} = \frac
{\sum_{g=2}^{\mathcal{T}}\sum_{t=2}^{\mathcal{T}}\mathbf{1}\{g\leq
t\}P(G=g) \cdot ATT(g,t)}{\sum_{g=2}^{\mathcal{T}}\sum_{t=2}^{\mathcal{T}}\mathbf{1}\{g\leq
t\}P(G=g)} \label{eqn:simplesum},%
\end{equation}
$(ii)$ the ``event-study-type'' causal parameter
\begin{equation}
\delta ^{es}\left( e\right) =\sum_{g=2}^{\mathcal{T}}\sum_{t=2}^{\mathcal{T}%
}1\left\{ t-g+1=e\right\} P\left( G_{g}=1|\text{Treated for $\ge$ e periods}\right) ATT\left(
g,t\right) ,  \label{eq:event-study}
\end{equation}
which provides the average treatment effect for units that have
been treated for $e$ periods, and $(iii)$ the average of $\delta ^{es}\left( e\right)$ over all possible (positive) values of $e$,
\begin{equation}
\delta^{e, avg}=\frac{1}{\mathcal{T}-1}\sum_{e=1}^{\mathcal{T}-1}\delta ^{es}\left( e\right). \label{eqn:dyn}%
\end{equation}
Note that all these weighted averages of the $ATT\left( g,t\right)$ are easy-to-interpret, less ``data hungry'' than the disaggregated $ATT\left( g,t\right)$, and can be use to summarize short, medium and long run effects of a given policy. In fact, one can show that (\ref{eq:ATS}) is equal to $\delta^{es}(e)$ with $e=1$.

The key challenge in estimating all these causal parameters of interest is to show that one can indeed nonparametrically point-identify all $ATT\left( g,t\right)$'s with $t \ge  g$. C\&S shows that one can bypass such challenges by imposing either the PTA \ref{ass:CS-PTA1} or \ref{ass:CS-PTA2}, though each assumption leads to a different estimand. More precisely, C\&S shows that, for $t \ge g$, $ATT\left( g,t\right)$ is nonparametrically identified by 
\begin{eqnarray}
ATT_{never}\left( g,t\right)  &=&\mathbb{E}\left[ \left.
Y_{t}-Y_{g-1}\right\vert G_{g}=1\right] -\mathbb{E}\left[ \left.
Y_{t}-Y_{g-1}\right\vert C=1\right] , \label{eq:att_never_pop}\\
ATT_{ny}\left( g,t\right)  &=&\mathbb{E}\left[ \left.
Y_{t}-Y_{g-1}\right\vert G_{g}=1\right] -\mathbb{E}\left[ \left.
Y_{t}-Y_{g-1}\right\vert D_{t}=0, G_{g}=0\right] ,
\label{eq:att_ny_pop}
\end{eqnarray}%
when one respectively imposes either the PTA \ref{ass:CS-PTA1} or \ref{ass:CS-PTA2}. These quantities can be straightforwardly estimated by
\begin{equation}
\widehat{ATT}_{never}\left( g,t\right) =\frac{n^{-1}\sum_{i=1}^{n}G_{ig}\left(
Y_{it}-Y_{ig-1}\right) }{n^{-1}\sum_{i=1}^{n}G_{ig}}-\frac{%
n^{-1}\sum_{i=1}^{n} C_{i} \left( Y_{it}-Y_{ig-1}\right) }{%
n^{-1}\sum_{i=1}^{n} C_{i} }
\label{eq:att_never}
\end{equation}%
and by $\widehat{ATT}_{ny}\left( g,t\right)$ as defined in (\ref{eq:att_ny}). 

With either $\widehat{ATT}_{never}\left( g,t\right)$ or $\widehat{ATT}_{ny}\left( g,t\right)$ on hand, one can then form the more aggregated parameters by replacing $\widehat{ATT}\left( g,t\right)$ by either of these estimators, and by replacing the weights by their natural (plug-in) estimators. For instance, depending on whether one imposes the parallel trends
assumption \ref{ass:CS-PTA1} or \ref{ass:CS-PTA2}, one can naturally
estimate $\delta^{es}\left( e\right)$  by 
\begin{eqnarray*}
\widehat{\delta }_{never}^{es}\left( e\right)  &=&\sum_{g=2}^{\mathcal{T}%
}\sum_{t=2}^{\mathcal{T}}1\left\{ t-g+1=e\right\} \widehat{w}(g;e) \widehat{ATT}_{never}\left( g,t\right) , \\
\widehat{\delta }_{ny}^{es}\left( e\right)  &=&\sum_{g=2}^{\mathcal{T}%
}\sum_{t=2}^{\mathcal{T}}1\left\{ t-g+1=e\right\} \widehat{w}(g;e) \widehat{ATT}_{ny}\left( g,t\right) ,
\end{eqnarray*}%
respectively, where $e\ge1$ and $\widehat{w}(g;e)$ is defined as in (\ref{eq:dyn-weights}). The aggregated estimators for $ATT^{simple}$ and for $\delta^{e,avg}$ are formed analogously.

C\&S derive the large sample properties of all these aforementioned estimators and propose bootstrap procedures to construct simultaneous
confidence bands for these treatment effect measures. They emphasize the practical importance of using simultaneous inference procedures when one estimates multiple parameters of interest (e.g., when one estimate $\delta^{es}(e)$ for multiple $e$'s), as failing to account for multiple testing usually lead to misleading inference.

We conclude this subsection by noting that S\&A is mainly interested in recovering the event-study-type parameter $\delta^{es}\left( e\right)$. More precisely, S\&A propose the following
interaction-weighted estimator for $\delta ^{es}\left( e\right)$ (see Section 4 of S\&A). In the
first-step, they use the linear two-way
fixed effects specification that interacts relative time indicators with
treatment group indicator:%
\begin{equation}
Y_{it}=\lambda _{i}+\lambda _{t}+\sum_{g=2}^{\mathcal{%
T}-1}\sum_{e \not=0}\delta
_{ge }\cdot G_{ig}1\left\{ t-G_{i}+1=e \right\} +v_{it}
\label{eq:IW}
\end{equation}%
on observations from $t=1,\dots ,\mathcal{T}-1$, where the last time period $%
\mathcal{T}$ is dropped in order to accommodate the case where there is no ``never treated'' group; if there is a never-treated group available, dropping data from time period $\mathcal{T}$ is unnecessary. S\&A shows that, under the PTA \ref{ass:dCdH-PTA}, the estimator $\widehat{\delta }_{ge }$ is consistent for $ATT\left( g,t \right) $, $t-g+1=e$. Here, it is important to emphasize that, when a ``never treated'' group is not available, only the units treated at the last time period are used as comparison units when computing $\widehat{\delta }_{ge }$, which differs from the C\&S proposed estimator (\ref{eq:att_ny}) that uses not-yet treated units as comparison units. When a ``never treated'' group is available, though, the interaction-weighted estimator $\widehat{\delta }_{ge }$ is equivalent to $\widehat{ATT}_{never}\left( g,t\right)$ once one maps event-time to calendar time (or vice-versa). 

Armed with $%
\widehat{\delta }_{ge }$, S\&A then propose to estimate $%
\delta ^{es}\left( e\right)$ by 
\begin{equation*}
\widehat{\delta }_{S\&A}^{es}\left( e\right) =\sum_{g=2}^{\mathcal{T}%
}\sum_{t=2}^{\mathcal{T}}1\left\{ t-g+1=e\right\} \widehat{w}(g;e) \widehat{\delta }_{gt},
\label{eq:AS}
\end{equation*}%
where $\widehat{w}(g;e)$ is defined as in (\ref{eq:dyn-weights}). S\&A establish the large sample properties of $%
\widehat{\delta }_{S\&A}^{es}\left( e\right) $ and provide valid pointwise inference
procedures for $\delta
^{es}\left( e\right)$.

\begin{remark}
\label{rm:pre-test} Many times one wishes to check for the existence of non-parallel pre-trends as a way to assess the credibility of the DID setup. We note that one can use $\delta^{es}_{never}(e)$ and/or $\delta^{es}_{ny}(e)$ with $e<0$ as estimators of pre-trends, though, when $e$ is negative one must replace the estimated weights $\widehat{w}(g;e)$ as defined as in (\ref{eq:dyn-weights}) with
\begin{equation*}
\widehat{w}(g;e-) \equiv \widehat{P}\left( G_{g}=1| \text{at least $|e|$ pre-treatment periods available}\right) =\frac{N_{g\cap \geq e-}}{%
N_{\geq e-}},
\end{equation*}%
where $N_{g\cap \geq e-}$ denotes the number of observations in group $g$ among
those units that have least $|e|$ pre-treatment time periods of data available, and $N_{\geq e-}$ is
the number of units who have least $|e|$ pre-treatment time periods of data available. Importantly, these event-study estimators avoid the pitfalls associated with using the dynamic TWFE to assess the credibility of parallel trends; see S\&A for a detailed discussion of this important issue.
\end{remark}


\section{Not all parallel trends assumptions are made equal\label{sec3}}
From the discussion in the previous section, it is clear that in DID designs with staggered treatment adoption one can make different parallel trends assumptions, and can estimate different parameters of interest. The discussion in Section \ref{sec2} also indicates that, despite their peculiarities, these different parameters of interest can be estimated using weighted averages of estimators for the $ATT(g,t)'s$. From this observation, one can reasonably argue that identifying (and estimating) the $ATT(g,t)'s$ from the data is the most challenging step of the analysis, and that the different PTAs help researchers to overcome it. Once this is done, constructing event-study-type estimators, for instance, becomes straightforward. 

In practice, however, researchers must choose and justify the use of a given PTA. In this section, we aim to highlight some practical consequences of adopting different versions of the PTA. In order to simplify the discussion, we exploit the stylized example introduced in Section \ref{sec2} whenever possible, and we implicitly impose Assumptions \ref{ass:iid}-\ref{ass:overlap}.

\subsection{Are these parallel trends testable?\label{sec:tests}}

Practitioners routinely use estimates of pre-treatment event-study coefficients to assess the credibility of an underlying PTA.\footnote{As stressed by S\&A, one should not use pre-treatment coefficients from TWFE event-study-type regressions to assess the credibility of the PTA as these coefficients can be contaminated with post-treatment effects. Using estimates of (\ref{eq:event-study}) with $e<0$, on the other hand, does not suffer from these pitfalls. Please refer to S\&A for a detailed discussion about these issues.} Can these tests for parallel pre-treatment trends be interpreted as \emph{direct} tests for the validity of the underlying PTA, or should these tests be interpreted as ``placebo/falsification'' type of tests?
With the help of the stylized example, we show that the answer depends on the chosen PTA.

Let us first consider the case of Assumption \ref{ass:dCdH-PTA}. As it is evident from (\ref{mom:3_3_1})-(\ref{mom:trends3}), the PTA \ref{ass:dCdH-PTA} imposes six linearly independent moment restrictions to recover three counterfactual parameters, $\alpha_{3,3}(0)$, $\alpha_{3,4}(0)$ and $ \alpha_{4,4}(0)$. That is, imposing the PTA \ref{ass:dCdH-PTA} leads to an overidentified system of equations and, consequently, we can directly test for the validity of the PTA \ref{ass:dCdH-PTA}. Indeed, it is easy to see that the PTA \ref{ass:dCdH-PTA} implies parallel pre-treatment trends across every group, see e.g., (\ref{mom:trends1})-(\ref{mom:trends3}), and such restrictions can be directly assessed from the data, for instance, by testing if pre-treatment event-study-type  estimates of (\ref{eq:event-study}) are all equal to zero. Thus, under the PTA   \ref{ass:dCdH-PTA}, non-zero pre-treatment event-study estimates should be interpreted as direct evidence against the identifying assumptions.\footnote{If Assumption \ref{ass:anticip} is violated, though, it is possible that violations from Assumption \ref{ass:anticip} ``offset" violations of Assumption \ref{ass:dCdH-PTA} and the test based on (\ref{mom:trends1})-(\ref{mom:trends3}) would not capture such violations. This should always be taken into account. In addition, failing to reject these tests should not be interpret as evidence in favor of the identifying assumptions, as it may be the case that the test lacks power to detect some non-trivial deviations from the null.}

A somewhat similar conclusion is also reached when one relies on the PTA \ref{ass:CS-PTA2}: (\ref{mom:33_ny})-(\ref{mom:44_ny}) suggest four linearly independent moment restrictions to recover three counterfactual parameters, which also leads to an overidentified system of equations. Recalling that (\ref{mom:33_ny})-(\ref{mom:44_ny}) is equivalent to (\ref{mom:3_3_1})-(\ref{mom:4_4_1}), we can then see from (\ref{mom:3_3_1})-(\ref{mom:3_3_2}) that the PTA \ref{ass:CS-PTA2} also imposes parallel pre-trends among ``never treated'' ($C=1$) and the ``later treated''($G_4=1$) from time $t=2$ to $t=3$, but does not restrict pre-trends of these two groups from $t=1$ to $t=2$, nor the pre-trends of the early treated group ($G_3=1$). Moving from the stylized example to the general case, we have that PTA \ref{ass:CS-PTA2} imposes parallel pre-treatment trends from $t=g_{min}-1$ to $t=\mathcal{T}$ for all groups except the first-treated group (who is treated at time $g_{min}$). Interestingly, because estimators of (\ref{eq:event-study}) with $e<0$ exploit \emph{all} pre-treatment trends (including the one for group $G_3=1$ in our stylized example), non-zero pre-treatment estimates can not, at least strictly speaking, be interpreted as direct tests of the identifying assumptions, but rather as placebo-type tests. Nonetheless, one can easily bypass this limitation by constructing alternative tests for the identifying assumptions. For instance, in the context of our stylized example, one can directly test if $\mathbb{E}\left[ Y_{3}-Y_{2}|C=1\right] = \mathbb{E}\left[ Y_{3}-Y_{2}|G_{4}=1\right]$ using a standard t-test. Rejecting the null hypothesis would provide direct evidence against the identifying assumptions.

Finally, note that the conclusion is very different when one imposes the PTA \ref{ass:CS-PTA1}: (\ref{mom:33})-(\ref{mom:44}) suggest we have a just-identified system of equations, implying that the PTA \ref{ass:CS-PTA1} can not be directly tested. Indeed, as we dicussed in Section \ref{sec:pta}, PTA \ref{ass:CS-PTA1} does not restrict pre-treatment trends, and, therefore, event-study estimates for pre-treatment periods provide, at best, placebo-type evidence against the PTA \ref{ass:CS-PTA1}.

The discussion above highlights that whether tests for parallel pre-treatment trends provide direct or indirect evidence against the invoked identifying assumptions crucially depends on the invoked PTA. This is very different from the case where treatment adoptions does not vary across time. In that case, tests for non-parallel pre-treatment trends \emph{always} provide only indirect evidence against the adopted design.

\subsection{Can the parallel trends assumption guide us on choosing a DID estimator?\label{sec:pta-estimators}}

In this section, we discuss some potential trade-offs one may face when adopting different PTAs, as they can lead to different DID estimators.

\subsubsection{What if we impose the PTA \protect\ref{ass:CS-PTA1}?}

The PTA \ref{ass:CS-PTA1} is the weakest PTA among the three we have considered so far as it does not impose any restriction on pre-treatment trends across groups. Given that this PTA leads to a just-identified system of equations, in situations where researchers are not willing to impose additional restrictions on the data,  $\widehat{AT}T_{never}\left( g,t\right) $ as defined in (\ref{eq:att_never}) is the \emph{only} suitable estimator for the $ATT\left( g,t\right)$ and their different functionals such as the event-study parameters (\ref{eq:event-study}). 

Of course, in order to rely on the PTA \ref{ass:CS-PTA1} and use the DID estimator (\ref{eq:att_never}), we must have a set of units that do not experience treatment in the time-window we want to analyze. When such a group of units is available but its relative size is small, inference procedures based on (\ref{eq:att_never}) may not be as precise as one wishes. However, it is important to stress that this potential ``loss of efficiency'' is a direct consequence of not exploiting restrictions on pre-treatment trends across groups. 

In practice, we foresee researchers taking into account this ``robustness'' versus ``efficiency'' trade-off when deciding if the PTA \ref{ass:CS-PTA1} is the most suitable for the given application. In situations where there is a ``reasonably large'' number of units that cannot be treated because of some application-specific institutional detail, we expect that the gains in robustness should dominate the potential gains in efficiency associated with using other PTAs and DID estimators. The same holds true if researchers are not comfortable with \emph{a priori} ruling out non-parallel pre-trends. In these cases, we foresee researchers favoring the PTA \ref{ass:CS-PTA1} and the DID estimator $\widehat{AT}T_{never}\left( g,t\right)$ over the other alternatives.

\subsubsection{What if we impose either the PTA \protect\ref{ass:dCdH-PTA} or the PTA \protect\ref{ass:CS-PTA2}?}

In many situations, a ``never-treated'' group is not available, implying that the PTA \ref{ass:CS-PTA1} does not provide any identifying restriction that can be used to estimate the $ATT(g,t)$'s. In other cases, the ``never-treated'' group may be ``too small'' to be of practical use, and/or researchers may be \emph{a priori} comfortable restricting pre-treatment trends and using those ``not-yet-treated'' units as valid comparison groups for those ``earlier-treated''. In such cases, researchers can then choose between the PTA \protect\ref{ass:dCdH-PTA} and the PTA \protect\ref{ass:CS-PTA2}. In both cases, though, they can use the the DID estimator $\widehat{AT}T_{ny}\left( g,t\right)$, as defined in (\ref{eq:att_ny}), to study policy effectiveness.

Although (\ref{eq:att_ny}) can be used under either PTA, we still recommend researchers to explicitly specify which PTA they are making for at least three reasons. First, being explicit about the identifying assumption adds transparency to the analysis, which is always desirable. Second, the interpretation of pre-tests based on event-study-type estimates can vary depending on the assumptions, as we discusses in Section \ref{sec:tests}. Third, the choice of the PTA has an important impact on what other estimators you could use instead of (\ref{eq:att_ny}). This is particularly important because (\ref{eq:att_ny}) does not fully exploit all the restrictions imposed by either PTA. Being aware of the exact PTA invoked allows researchers to adopt an alternative estimation procedure that fully exploits all these moment restrictions, resulting in estimators that are more efficient than (\ref{eq:att_ny}). Here, we stress that the gains in efficiency will vary depending on the underlying PTA used, as the PTA \ref{ass:dCdH-PTA} imposes more restrictions on the data than PTA \ref{ass:CS-PTA2}. 

Before describing how one can exploit these additional moment restrictions to form a more efficient DID estimator, it is worth describing situations where researchers may favor either the PTA \ref{ass:dCdH-PTA} or the PTA \ref{ass:CS-PTA2}. Recall that the main difference between PTA \protect\ref{ass:dCdH-PTA} and PTA \protect\ref{ass:CS-PTA2} is that the former imposes parallel pre-treatment trends across all groups and all time periods, whereas the latter only restricts pre-treatment trends since the time the first group is treated. These differences can be meaningful in applications where data on multiple time periods before the first group of units is treated are available, and the economic environment in these ``early-periods'' were potentially different from the ``later-periods''. In these cases, the outcome of the different groups may evolve in a non-parallel manner during ``early-periods'' periods, perhaps because the groups were exposed to different shocks, but these non-parallel trends become less of a concern as time pass by. In such cases, we expect researchers to favor the PTA \protect\ref{ass:CS-PTA2} over the PTA \protect\ref{ass:dCdH-PTA}. In other situations, though, researchers may prefer to impose the PTA \protect\ref{ass:dCdH-PTA}, allowing them to enjoy some potential gains in efficiency if they  use estimators that exploit the additional restrictions imposed by the PTA \ref{ass:dCdH-PTA} when compared to the PTA \ref{ass:CS-PTA2}. Again, the ``robustness'' versus ``efficiency'' trade-off should be taken into account when deciding which PTA is more appropriate for the specific application.

\section{Using GMM to estimate DID parameters\label{sec:GMM}}
As we described in Section \ref{sec:pta-estimators}, in situations where researchers are comfortable with imposing either the PTA \ref{ass:dCdH-PTA} or the PTA \ref{ass:CS-PTA2}, the DID estimator (\ref{eq:att_ny}) is not efficient, as it does not fully exploit all the restrictions implied by these PTAs. In this section, we describe how one can exploit all the restrictions implied by the identifying assumptions to form efficient DID estimators by casting the problem into the familiar GMM framework \citep{Hansen1982}.  In what follows, we provide a step-by-step description of how one can form these efficient GMM DID estimators. To avoid repetition, we focus on the case where researchers impose the PTA \ref{ass:CS-PTA2}; the implementation based on the PTA \ref{ass:dCdH-PTA} is completely analogous. 

The key to implement the GMM is to list \emph{all} moment restrictions we are imposing to recover the $ATT(g,t)$'s, which involves not only the moment restrictions implied by the PTA \ref{ass:CS-PTA2}, but also the  \emph{observational restrictions} that, for all $t\ge g$,
$\alpha _{g,t}\left( 1\right)  \equiv \mathbb{E}\left[ Y_{t}\left( 1\right)|G_{g}=1\right] =\mathbb{E}\left[ Y_{t}|G_{g}=1\right],$  
$\alpha _{g}^{prop} \equiv \mathbb{E}\left[ G_{g}\right]$,   
$\alpha _{C}^{prop} \equiv \mathbb{E}\left[ C\right]$.
We can then use these ``augmented'' moment restrictions (consisting of observational restriction and all the moment restrictions implied by the PTA) to efficiently estimate all the unknown parameters involved in our problem by following \cite{Hansen1982}. 

To gain more intuition on how to implement the efficient GMM, we turn of attention to our stylized example. In this case, the unknown parameters consist of $\alpha \equiv \left( \alpha _{3,3}\left( 1\right) ,\alpha_{3,3}\left( 0\right) ,\alpha _{3,4}\left( 1\right) ,\alpha _{3,4}\left(0\right) ,\alpha _{4,4}\left( 1\right) ,\alpha_{4,4}\left( 0\right) ,\alpha_{C}^{prop},\alpha _{3}^{prop}\text{,}\alpha _{4}^{prop}\right) ^{\prime }$, which can be efficiently estimated by 
\begin{equation}
\widehat{\alpha }^{gmm}=\arg \min_{\alpha \in \Theta }\bar{g}_{\alpha
}\left( W\right) ^{\prime }\widehat{\Sigma }_{\check{\alpha},gmm}^{-1}\bar{g}%
_{\alpha }\left( W\right) ,  \label{eq:gmm}
\end{equation}%
where $\bar{g}_{\alpha }\left( W\right) $ is the sample average of the
augmented moment conditions, $n^{-1}\sum_{i=1}^{n}g_{\alpha }\left(W_{i}\right)$,
with $g_{a}\left( W_{i}\right) $ combining all (linearly independent) moment conditions\footnote{See Web Appendix A for the details about $g_{a}\left( W_{i}\right)$.},and 
\begin{equation*}
\widehat{\Sigma }_{\check{\alpha},gmm}=\frac{1}{n}\sum_{i=1}^{n}g_{\check{%
\alpha}}\left( W_{i}\right) g_{\check{\alpha}}\left( W_{i}\right) ^{\prime },
\end{equation*}%
 $\check{\alpha}$ being a preliminary consistent estimator for $\alpha $,
say the minimizer of (\ref{eq:gmm}) with $\widehat{\Sigma }_{\check{\alpha}%
,gmm}$ replaced by the identity matrix.

With $\widehat{\alpha }^{gmm}$, one can then efficiently estimate the parameters of interest: $ATT\left(
3,3\right) $, $ATT\left( 3,4\right) $ and $ATT\left( 4,4\right) $ by%
\begin{equation}
\widehat{ATT}_{gmm}\left( 
\begin{array}{c}
3,3 \\ 
3,4 \\ 
4,4%
\end{array}%
\right) =\left( 
\begin{array}{c}
\widehat{\alpha }_{3,3}^{gmm}\left( 1\right) -\widehat{\alpha }%
_{3,3}^{gmm}\left( 0\right)  \\ 
\widehat{\alpha }_{3,4}^{gmm}\left( 1\right) -\widehat{\alpha }%
_{3,4}^{gmm}\left( 0\right)  \\ 
\widehat{\alpha }_{4,4}^{gmm}\left( 1\right) -\widehat{\alpha }%
_{4,4}^{gmm}\left( 0\right) 
\end{array}%
\right) .  \label{eq:att_gmm_est}
\end{equation}

In what follows, we establish the asymptotic properties of $\widehat{\alpha }%
^{gmm}$. The asymptotic properties of $\widehat{ATT}_{gmm}\left( g,t\right) $
 follow directly from the delta method. Let $\Delta Y_{t}=Y_{t}-Y_{t-1}$. Define
$\Sigma _{\alpha ,gmm}$ as the probability limit of $\widehat{%
\Sigma }_{\tilde{\alpha},gmm}$, and
\begin{eqnarray*}
\Psi  &=&\mathbb{E}\left[ \frac{\partial g_{\alpha }\left( W\right) }{%
\partial \alpha }\right]
\end{eqnarray*}%
Let the vector of scores associated with the efficient GMM estimator be defined as
\begin{equation*}
\phi_{\alpha }^{gmm}\left( W_{i}\right) =-\left( \Psi ^{\prime }\Sigma
_{\alpha ,gmm}^{-1}\Psi \right) ^{-1}\Psi ^{\prime }\Sigma _{\alpha
,gmm}^{-1}\cdot g_{\alpha }\left( W_{i}\right) .  \label{eq:lin.rep.gmm}
\end{equation*}

\begin{proposition}
\label{Thm:1}Assume that all random variables have finite second moments, $\Sigma _{\alpha,gmm}$ is positive definite, and that Assumptions \ref{ass:iid}-\ref{ass:overlap} hold. Then, when the parallel trends assumption \ref{ass:CS-PTA2} holds, we have that:

$\left( i\right) $ As $n\rightarrow \infty $, 
\begin{eqnarray*}
\sqrt{n}\left( \widehat{\alpha }^{gmm}-\alpha \right)  &=&\frac{1}{\sqrt{n}}%
\sum_{i=1}^{n}\phi_{\alpha }^{gmm}\left( W_{i}\right) +o_{p}\left(
1\right)   \label{eq:lin.rep_gmm} \\
&&\overset{d}{\rightarrow }N\left( 0,\left( \Psi ^{\prime }\Sigma _{\alpha
,gmm}^{-1}\Psi \right) ^{-1}\right) .  \notag
\end{eqnarray*}%
$\left( ii\right) $ The GMM estimator $\widehat{\alpha }^{gmm}$ is
semiparametrically efficient.
\end{proposition}
\begin{proof}
Proof is presented in the Web Appendix C.
\end{proof}

Proposition \ref{Thm:1} has important practical implications, which we illustrate in the context of our stylized example. First, and
perhaps most important, it implies that 
\begin{equation*}
\sqrt{n}\left( \widehat{ATT}_{gmm}-ATT\right) \left( 
\begin{array}{c}
3,3 \\ 
3,4 \\ 
4,4%
\end{array}%
\right) \overset{d}{\rightarrow }N\left( 0,\Omega \right) 
\label{eq:ATT_gmm}
\end{equation*}%
with $\Omega =A\left( \Psi ^{\prime }\Sigma _{\alpha ,gmm}^{-1}\Psi \right)
^{-1}A^{\prime }$, A is a ``selection matrix''\footnote{See Web Appendix A for its formal definition.}
and that $\Omega$ is equal to the semiparametric efficiency bound for $ATT\left(
g,t\right) $ under the PTA \ref{ass:CS-PTA2}. As so, $\widehat{ATT}_{gmm}$
exploits all available information in the data to estimate the $ATT\left(
g,t\right) $'s, which, in general, translates to tighter confidence
intervals. In fact, under the PTA \ref{ass:CS-PTA2}, the GMM DID estimator $\widehat{ATT}_{gmm}$ $\left( g,t\right)$ is, in general, more efficient
than $\widehat{ATT}_{ny}\left( g,t\right)$ or $\widehat{ATT}_{never}\left(g,t\right) $ as defined in (\ref{eq:att_ny}), and (\ref{eq:att_never}), respectively, or those based on the “interaction-weighted'' regression (\ref{eq:IW}). This is a main advantage of the GMM DID estimator when compared to the other available estimators.

A second implication of Proposition \ref{Thm:1} is that, given that we have
an \emph{overidentified} system of equations, one can directly use the
Sargan-Hansen J-test as a test for the validity of the PTA \ref{ass:CS-PTA2}. More precisely, under the null hypothesis that the PTA \ref{ass:CS-PTA2} is true, 
\begin{equation*}
J=n\cdot \left( \bar{g}_{\widehat{\alpha }^{gmm}}\left( W\right) ^{\prime }~%
\widehat{\Sigma }_{\widehat{\alpha }^{gmm},gmm}^{-1}~\bar{g}_{\widehat{%
\alpha }^{gmm}}\left( W\right) \right) \overset{d}{\rightarrow }\chi
_{10-9}^{2}\text{ as }n\rightarrow \infty .  \label{eq:J-test}
\end{equation*}%
If the PTA \ref{ass:CS-PTA2} holds, any deviation of $J$ from zero should
be within the range of sampling error, whereas if the PTA \ref{ass:CS-PTA2} is violated, $J$ should be ``large.'' Thus, the Sargan-Hanen J-test can be useful for detecting violations of the PTA \ref{ass:CS-PTA2}.

At this moment, one may wonder about the situations where one may favor the GMM DID estimator (\ref{eq:att_gmm_est}) over the simpler DID estimator (\ref{eq:att_ny}). Given that (\ref{eq:att_gmm_est}) is more efficient than and as-robust-as (\ref{eq:att_ny}), the only obstacle we see for its widespread adoption is its implementation: whenever the number of treatment groups and/or time periods available are large, the number of moment conditions needed to be considered into the efficient GMM can be fairly large --- in our application, for example, where we have 16 treatment groups and 33 time periods, the GMM involves 780 moments with \emph{195 overidentification restrictions}, whereas sample size (state-year pairs) is equal to 759. In such cases, we expect researchers to favor the simpler, but inefficient DID estimator (\ref{eq:att_ny}). However, when the number of groups and/or time periods is moderate such that implementation of the efficient GMM is not challenging, we would recommend using it.

\section{A simpler and more robust DID estimator\label{sec:Alternative}}

As highlighted in the previous section, the main attractive feature of using the GMM estimation procedure is that it leads to efficient estimators that fully exploit all the available information compatible with the underlying identifying assumptions. On the other hand, the implementation of such GMM DID estimator is not always straightforward.

In this section, we describe an alternative DID estimator for the $ATT(g,t)$. Although the estimator is usually less efficient than the GMM, it is (i) easy to compute, (ii) does not require the existence of a ``never-treated'' group, (iii) exploits more data than (\ref{eq:att_ny}), and (iv) does not explicitly restrict pre-trends as it relies on a ``weaker'' PTA described below.

\begin{assumption}[``Weaker'' Parallel trends assumption based on \textquotedblleft
not-yet treated\textquotedblright\ units]
\label{ass:W-PTA} For all $g,t =2,\ldots ,\mathcal{T}$, such that $t\geq g$, 
\begin{equation*}
\mathbb{E}\left[ \left. Y_{t}\left( 0\right) -Y_{t-1}\left( 0\right)
\right\vert G_{g}=1\right] =\mathbb{E}\left[ \left. Y_{t}\left( 0\right)
-Y_{t-1}\left( 0\right) \right\vert D_{t}=0\right] .
\end{equation*}
\end{assumption}

The PTA \ref{ass:W-PTA} imposes that the evolution of the outcome at time $t$ among those units that have not yet experienced treatment by time $t$ can help us identify the $ATT(g,t)$'s. Unlike PTA \ref{ass:CS-PTA2}, it does not impose that \emph{every individual} not-yet-treated group can be used as a comparison group, which, in turn, suggests that the $ATT(g,t)$, $t\ge g$ are nonparametrically \emph{just-identified}. We formalize this result in the next proposition. Let $\Delta Y_t = Y_{t}-Y_{t-1}$ denote the first-difference of $Y_{t}$.

\begin{proposition}
\label{Thm:identifi} Assume that Assumptions \ref{ass:iid}-\ref{ass:overlap} hold. Then, when the parallel trends assumption \ref{ass:W-PTA} holds, it follows that, for $2\le g \le t \le \mathcal{T}$, $ATT(g,t) = ATT_{ny+}\left( g,t\right)$, where
\begin{equation}
ATT_{ny+}\left( g,t\right) \equiv \mathbb{E}\left[ \left.
Y_{t}-Y_{g-1}\right\vert G_{g}=1\right] - \left(\sum_{s=g}^{t}\mathbb{E}\left[ \left.
\Delta Y_{s}\right\vert D_{s}=0, G_g=0\right] \right). \label{eq:att_ny_weak}
\end{equation}%
\end{proposition}
\begin{proof}
Proof is presented in the Web Appendix C.
\end{proof}

To better grasp how the PTA \ref{ass:W-PTA} allows us to use $ATT_{ny+}\left( g,t\right)$ as an estimand for the $ATT(g,t)$, $t\ge g$, it is illustrative to go back to our stylized example. In this specific context, we have that, under Assumptions \ref{ass:iid}-\ref{ass:overlap}, the PTA \ref{ass:W-PTA} is equivalent to the following restrictions:
\begin{eqnarray}
\alpha _{3,3}\left( 0\right)  &=&\mathbb{E}\left[ Y_{3}-Y_{2}|D_{3}=0\right] +%
\mathbb{E}\left[ Y_{2}|G_{3}=1\right] ,  \label{mom:33_W} \\
\alpha _{3,4}\left( 0\right)  &=&\mathbb{E}\left[ Y_{4}-Y_{3}|D_{4}=0\right]
+\alpha _{3,3}\left( 0\right) ,  \label{mom:34_W} \\
\alpha _{4,4}\left( 0\right)  &=&\mathbb{E}\left[ Y_{4}-Y_{3}|D_{4}=0\right] +%
\mathbb{E}\left[ Y_{3}|G_{4}=1\right] .  \label{mom:44_W}
\end{eqnarray}%

By listing these restrictions we can now see how we get $ATT_{ny+}\left( g,t\right)$, as defined in (\ref{eq:att_ny_weak}). First, from (\ref{mom:33_W}) and (\ref{mom:44_W}), it follows that, when $g=t$, $\alpha_{g,t}(0)$ can be explicitly written as functionals of observable data (and not potential outcomes). As so, $ATT(g,t)$ is identified by (\ref{eq:att_ny_weak}). Interestingly, in this case with $g=t$, (\ref{eq:att_ny_weak}) reduces to $ATT_{ny}(g,t)$, as defined in (\ref{eq:att_ny_pop}). When one moves away from the ``instantaneous average treatment effects'', though, these two estimands differ. Indeed, by exploiting the moment restrictions (\ref{mom:33_W}) and (\ref{mom:34_W}), we can see that the $ATT(3,4)$ is nonparametrically identified by 
\begin{equation*}
ATT_{ny+}\left( 3,4\right) = \mathbb{E}\left[ \left.
Y_{4}-Y_{2}\right\vert G_{3}=1\right] - \left(\mathbb{E}\left[ \left.
Y_{3}-Y_{2}\right\vert D_{3}=0\right] + \mathbb{E}\left[ \left.
Y_{4}-Y_{3}\right\vert D_{4}=0\right] \right),
\end{equation*}
which, of course, is a special case of (\ref{eq:att_ny_weak}). Here, we stress that  $ATT_{ny+}\left( 3,4\right)$ uses data from all groups, $G_3=1$, $G_4=1$, and $C=1$, whereas $ATT_{ny}\left( 3,4\right)$ only uses data from $G_3=1$ and $C=1$. Hence, one may expect that estimators based on (\ref{eq:att_ny_weak}) to be more precise than (\ref{eq:att_ny}) because they utilize more data. Furthermore, because the PTA \ref{ass:W-PTA} does not restrict pre-trends, i.e., it does not impose that $\mathbb{E}\left[ \left.
Y_{3}-Y_{2}\right\vert D_{3}=0\right] = \mathbb{E}\left[ \left.
Y_{3}-Y_{2}\right\vert D_{4}=0\right]$ as implied by (\ref{mom:33_ny}) and (\ref{mom:33_new_ny}), one can also expect additional gains in ``robustness'' by exploiting (\ref{eq:att_ny_weak})  instead of (\ref{eq:att_ny_pop}).

Next, we discuss how one can exploit Proposition \ref{Thm:identifi} to estimate the $ATT(g,t)$s. Here, the most natural way to proceed is to use the sample analogue of (\ref{eq:att_ny_weak}):
\begin{equation}
\widehat{ATT}_{ny+}\left( g,t\right) =\frac{n^{-1}\sum_{i=1}^{n}G_{ig}\left(
Y_{it}-Y_{ig-1}\right) }{n^{-1}\sum_{i=1}^{n}G_{it}}- \sum_{s=g}^{t}\left(\frac{
n^{-1}\sum_{i=1}^{n}\left( 1-D_{is}\right) \left( 1-G_{ig}\right) \Delta Y_{is}}{%
n^{-1}\sum_{i=1}^{n}\left( 1-D_{is}\right) \left( 1-G_{ig}\right)}\right).
\label{eq:att_ny+}
\end{equation}%

Note that (\ref{eq:att_ny+}) is very easy to compute as it only involves combinations of sample means. Next, we show that these DID estimators also enjoy good asymptotic properties. More precisely, we prove they are $\sqrt n$-consistent and establish their \emph{joint} asymptotic distribution. Before we present the results, we need to introduce some additional notation. For each $(g,t)$-pair, let $\phi_{ny+}\left( W_{i} ;g,t\right)$ be the influence function of $\widehat{ATT}_{ny+}\left( g,t\right)$,
\begin{multline*}
\phi _{ny+}\left(W_{i}; g,t\right) = \left( \frac{G_{ig}}{%
\mathbb{E}[G_{g}]}\left( \left( Y_{it}-Y_{ig-1}\right) -\frac{\mathbb{E}%
[G_{g}\cdot \left( Y_{t}-Y_{g-1}\right) }{\mathbb{E}[G_{g}]}\right) \right. 
\\
-\left. \sum_{s=g}^{t}\frac{\left( 1-D_{is}\right) \left( 1-G_{ig}\right) }{%
\mathbb{E}[\left( 1-D_{s}\right) \left( 1-G_{g}\right) ]}\left( \Delta
Y_{is}-\frac{\mathbb{E}[\left( 1-D_{s}\right) \left( 1-G_{g}\right) \cdot
\Delta Y_{s}]}{\mathbb{E}[\left( 1-D_{s}\right) \left( 1-G_{g}\right) ]}%
\right) \right).
\end{multline*}
Finally, let $\widehat{ATT}_{ny+}(t \ge g)$ and $ATT_(t \ge g)$ denote the vector of $\widehat{ATT}_{ny+}(g,t)$ and $ATT(g,t)$, respectively, for all $g,t=2,\dots,\mathcal{T}$ with $t \ge g$. Analogously, let $\Phi_{ny+}\left(W_{i}; t \ge g\right)$ denote the collection of $\phi _{ny+}\left(W_{i}; g,t\right)$ across all periods t and groups g such that $t \ge g$. \footnote{We restrict our attention to $t \ge g$ just because these are the post-treatment periods, which presumably are the periods of main interest for the analysis. However, our results naturally extend to the case where one consider all possible $g,t$'s, with the caveat that $ATT_{ny+}(g,t)$ may differ from $ATT(g,t)$ for $t < g$, as the PTA \ref{ass:W-PTA} does not explicit restrict pre-trends.}

\begin{proposition}
\label{Thm:Asy_prop} Assume that Assumptions \ref{ass:iid}-\ref{ass:overlap} and Assumption \ref{ass:W-PTA} hold. Then, as $n\rightarrow \infty $, 
\begin{equation}
\sqrt{n}\left( \widehat{ATT}_{ny+} - ATT \right)\left( g,t\right)  =\frac{1}{\sqrt{n}}%
\sum_{i=1}^{n}\phi_{ny+}\left(W_{i}; g,t\right) +o_{p}\left(
1\right).   \label{eq:lin.rep} 
\end{equation}%
Furthermore, 
\begin{equation}
\label{eq:CLT}
\sqrt{n}\left( \widehat{ATT}_{ny+}(t\ge g) - {ATT}(t\ge g) \right) \overset{d}{\rightarrow }N\left( 0,V\right), 
\end{equation}%
with $V = \mathbb{E} \left(\Phi_{ny+}\left(W; t\ge g\right) \Phi_{ny+}\left(W; t\ge g\right)'\right)$.
\end{proposition}
\begin{proof}
Proof is presented in the Web Appendix C.
\end{proof}

Proposition \ref{Thm:Asy_prop} provides the influence function for estimating the vector of group-time average treatment effects, as well as its limiting distribution.  Interestingly, Proposition \ref{Thm:Asy_prop} suggests two potential paths to conduct
inference for the $ATT\left( g,t\right)$'s. The first and perhaps more
standard approach is to use the analogy principle and directly estimate $V$, which leads directly to standard errors and pointwise confidence intervals. However, it is worth stressing that when one is interested in making inference about \emph{multiple} $ATT\left( g,t\right) $'s, inference procedures based on this standard approach such as those based
on traditional t-tests and/or individual confidence intervals are usually
inappropriate as they do not account for the fact that one is (implicitly)
conducting \emph{multiple hypotheses testing}. As a direct consequence,
significant treatment effects may emerge simply by chance, even when all $%
ATT\left( g,t\right) $'s are equal to zero, see, e.g., \cite{Romano2005}, \cite%
{Anderson2008} and section 8 of \cite{Romano2010}.

An alternative path to conduct asymptotically valid inference for multiple
parameters of interest that is robust against the multiple-testing problem
is to leverage the asymptotic linear representation (\ref{eq:lin.rep}) to
construct computationally-simple bootstrapped \emph{simultaneous}
confidence intervals for multiple $ATT\left( g,t\right) $. The idea of this
bootstrap procedure is fairly simple, and each bootstrap iteration simply
amounts to \textquotedblleft perturbing\textquotedblright\ the asymptotic linear representation of the $\widehat{ATT}_{ny+}\left( g,t\right)$'s by a random weight $V$, and it does not require re-estimating the $ATT(g,t)$'s at each bootstrap draw. In Web Appendix B, we provide a step-by-step description of how one can implement such a procedure.  

\begin{remark}\label{rem:new-pre-trends-est}
It is worth stressing that the $ATT_{ny+}\left( g,t\right)$ estimand defined in (\ref{eq:att_ny_weak}) is only suitable for post-treatment periods, i.e., for $t \ge g$. Hence, in contrast to (\ref{eq:att_never_pop}) and (\ref{eq:att_ny_pop}), we can not fix the estimand to analyze pre-treatment periods $t < g$. To address this issue,  we suggest using the estimand 
\begin{equation}
ATT_{ny+}^{pre}\left( g,t\right) \equiv \mathbb{E}\left[ \left.
\Delta Y_{t}\right\vert G_{g}=1\right] - \mathbb{E}\left[ \left.
\Delta Y_{t}\right\vert D_{t}=0, G_g=0\right],~~~for~~t < g, \label{eq:att_ny_weak_pre}
\end{equation}%
which should be equal to zero under Assumptions \ref{ass:iid}-\ref{ass:overlap} and a stronger version of the PTA \ref{ass:W-PTA} that holds for both pre and post-treatment periods (and not only for post-treatment periods as PTA \ref{ass:W-PTA}). One can then use estimates of (\ref{eq:att_ny_weak_pre}) to provide indirect evidence for the PTA \ref{ass:W-PTA}, as PTA \ref{ass:W-PTA} cannot be directly tested.
We stress that (\ref{eq:att_ny_weak_pre}) should not be directly compared with (\ref{eq:att_never_pop}) and (\ref{eq:att_ny_pop}) for $t < g$ as (\ref{eq:att_ny_weak_pre}) measures ``local deviations'' (from time $t-1$ to $t$)  of a zero pre-treatment trends conditions, whereas (\ref{eq:att_never_pop}) and (\ref{eq:att_ny_pop}) capture ``cumulative deviations'' (from time $t$ until $g-1$) of zero pre-treatment trends conditions.
\end{remark}

\begin{remark}\label{rem:event-study}
It is straightforward to build on $\widehat{ATT}_{ny+}\left( g,t\right)$ to construct event-study estimators for $\delta^{es}(e)$ as defined in (\ref{eq:event-study}). Following the same steps described in Section \ref{sec:parameters}, a natural estimator for $\delta^{es}(e)$ is \begin{eqnarray*}
\widehat{\delta }_{ny+}^{es}\left( e\right)  &=&\sum_{g=2}^{\mathcal{T}%
}\sum_{t=2}^{\mathcal{T}}1\left\{ t-g+1=e\right\} \widehat{w}(g;e) \widehat{ATT}_{ny+}\left( g,t\right),
\end{eqnarray*}
where the weights $\widehat{w}(g;e)$ are as defined in (\ref{eq:dyn-weights}). By building on Theorem \ref{Thm:Asy_prop} and the fact that the weights admit an asymptotic linear representation, it is easy to show that $\widehat{\delta }_{ny+}^{es}\left( e\right)$ is consistent and asymptotically normal; see, e.g. C\&S.
\end{remark}

We conclude this section by highlighting situations where we foresee (\ref{eq:att_ny+}) being favored over the other available DID estimators. First, we envision researchers favoring (\ref{eq:att_ny+}) over (\ref{eq:att_ny}) in situations where they are not comfortable explicitly restricting pre-trends and/or when they want to use data from all groups to estimate the $ATT(g,t)$'s. This can be particularly relevant when one wants to conduct cluster-robust inference when only a moderate number of clusters is available. We also expect researchers to favor (\ref{eq:att_ny+}) over the efficient GMM estimator when implementation is challenging. In this case, we expect that (\ref{eq:att_ny+})'s ``easiness-to-use'' would dominate the potential efficiency gains of using GMM. Finally, we expect researcher to favor (\ref{eq:att_ny+}) over (\ref{eq:att_never}) when a ``never-treated'' group is relatively small, though we stress that these two estimators rely on non-nested PTAs.

\begin{remark}
Given that different estimators (and PTAs) have different implications for robustness and efficiency, it may be tempting to 
\textquotedblleft specific-to-general\textquotedblright\ specification
search: start the analysis considering estimators that rely on ``stronger'' assumptions and then test the validity of these assumptions;  in case one does not reject them, one stop and uses the ``more efficient'' estimators, but in case one rejects the invoked PTA, one then chooses a ``more robust'' but ``less efficient'' DID estimator. Although fairly intuitive, the aforementioned strategy is dangerous and \emph{should not be used} in practice since this specification search is based on multiple-testing procedure, and, as so, inference procedures that treat the \textquotedblleft final\textquotedblright\ estimator (or the \textquotedblleft winner\textquotedblright ) as  \textquotedblleft true\textquotedblright\ can be severely distorted, see, e.g. \cite{Roth2019} for detailed discussion of this issue. Hence, we argue that researchers should select the PTA taking into account the \textquotedblleft robustness\textquotedblright\ versus \textquotedblleft efficiency\textquotedblright\ trade-off, and that these considerations should be done based on external, context-specific information, and not on pre-tests. 
\end{remark}

\section{The effect of enforcing the Clean Water Act\label{sec:Application}}

To illustrate the inherent trade-offs described above, we replicate Katherine Grooms' (2015) analysis of the transition from federal to state management of the Clean Water Act (CWA). Environmental policy mandated at the federal level is often implemented at the state level. Yet, there exists variation in the level of enforcement across states. \citet{grooms2015} exploits the staggered timing of the transfer from federal to state monitoring and enforcement of the CWA. Using TWFE specifications akin to (\ref{eq:TWFE-1}) and (\ref{eq:TWFE2}), she finds that state-level prevalence of corruption plays an important role in the enforcement and compliance of environmental regulation after transitioning to state control. 

We begin by describing the data, and then we discuss the practical relevance of the key assumptions and specifications we use for the analysis given our context. Finally, we show the baseline and corruption-specific results for both the TWFE specification and the new DID estimators that rely on the different PTAs discussed above. Finally, we discuss the implications of the findings and the importance of choosing an appropriate PTA.

\subsection{Data}
We follow the data construction from \citet{grooms2015} as closely as possible. Table~D.1 in Web Appendix D replicates key summary statistics from \citet{grooms2015} and provides additional detail on data sources and construction. As described further in Web Appendix D, we follow \citet{grooms2015} to construct a measure of the fraction of total facilities with at least one inspection, violation, or enforcement action in a state and year.

\begin{figure}[htb]
\centering
\caption{Timing of State Authorization} \label{fig:year}
\includegraphics[width=.75\linewidth, keepaspectratio]{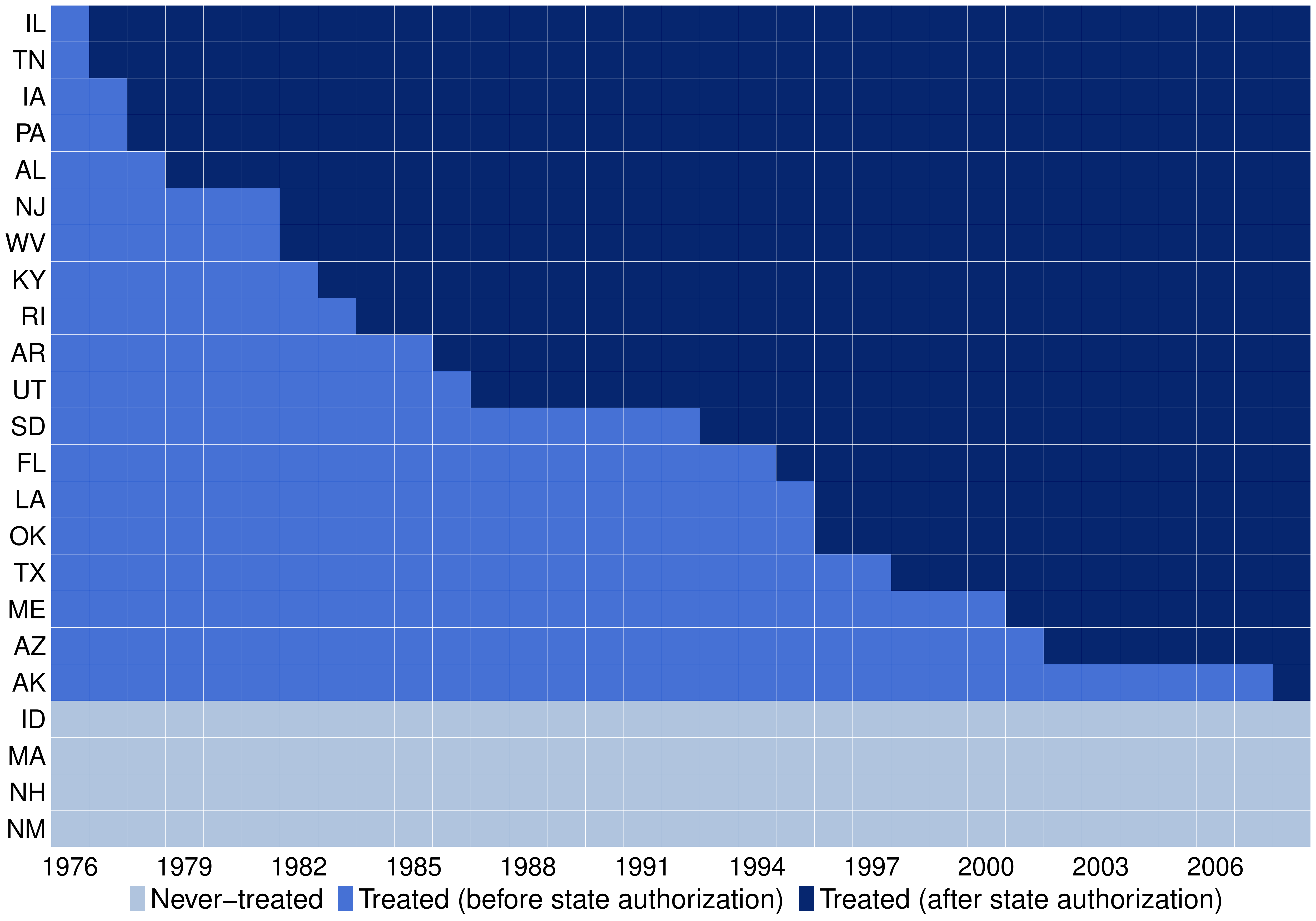} \\
\caption*{\footnotesize Notes: Shows the timing treatment adoption, where treatment is defined as the year in which the state was authorized to administer individual NPDES permits.}
\end{figure}

The timing of state authorization is distributed fairly evenly throughout our sample period, with the exception that 27 states received authorization prior to the sample period, between 1973-1975.\footnote{ Figure~D.1 in the Web Appendix D shows the distribution of the timing of state authorization across years. As many states receive authorization for the first four phases in the same year, we define the year of authorization as the year in which the state was authorized to perform the first phase of the program, administering individual NPDES permits.} Given that neither the data nor the parallel trends assumptions for $Y_t(0)$ provide information to identify the average treatment effect for these ``always treated'' states, these states are dropped from the analysis. Figure~\ref{fig:year} highlights the year that each of the remaining 23 states started treatment, i.e., the year in which the state was authorized to administer individual NPDES permits. The bottom four states are what we call the ``never-treated'' units, i.e., the states that remain unauthorized to administer individual NPDES permits through the entire sample period.\footnote{As of 2008, four states remained unauthorized to administer individual NPDES. Idaho received authorization in 2018, outside of the sample period used here to be consistent with \citet{grooms2015}.} Figure~\ref{fig:year} also allows one to visualize which states form each treatment group (those states whose colors turn to dark blue in the same year), and who the ``not-yet-treated'' states are at any point in time (those units that are colored light-blue in a given year).

\begin{figure}
    \centering
    \caption{Corrupt and non-corrupt states}
    \label{fig:corrupt}
    \includegraphics[width=.8\linewidth, keepaspectratio]{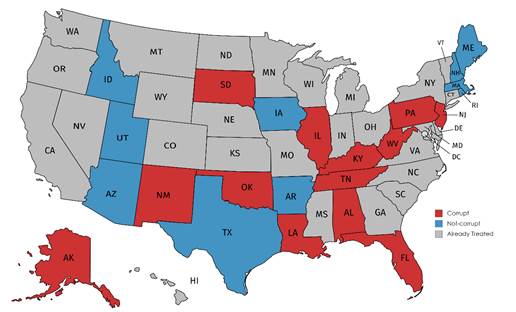}
    \caption*{\footnotesize Notes: Corrupt states, shown in red, are those above the median of average convictions per capita across all years. Non-corrupt states are shown in blue. Grey states are ``already treated'' prior to the sample window and are not included in the analysis.}
\end{figure}

Finally, we follow \cite{grooms2015} in defining states with above median federal public corruption convictions across all years as ``corrupt'' states.\footnote{See Web Appendix D for additional detail.} Figure \ref{fig:corrupt} shows corrupt states in red and non-corrupt states in blue. ``Always-treated'' states are shown in grey. Based on this measure, ``corrupt'' states are mostly from the mid-Atlantic and southern regions, while ``non-corrupt'' states tend to appear in New England and the west.

\subsection{Specifications and assumptions}
Like \cite{grooms2015}, the starting point of our exercise is to examine the impact of authorization on compliance outcomes --- for the sake of brevity, we focus on violation rates, though results for inspection rate and enforcement rate are available upon request.\footnote{Overall, we find essentially zero effect on these other outcomes, regardless of the PTA and model specification used. This is in line with the results in \cite{grooms2015}.} 

Since we are particularly interested in treatment effect dynamics, we estimate event-study-type parameters using four different procedures. First, we replicate the dynamic TWFE specification from \cite{grooms2015}. The exact specification we use is the following:
\begin{equation}
Y_{it}=\lambda _{i}+\lambda _{t}+\sum_{e =-30, e\not=0}^{32}\beta _{e }1\left\{
t-G_{i}+1=e \right\} +v_{it},
\label{eq:event1-app}
\end{equation}%
which includes 30 treatment lead indicators (all the indicators associated with $\beta_{e}$ with $e<0$) and 32 treatment lag indicators (all the indicators associated with $\beta_{e}$ with $e>0$). We follow \cite{Borusyak2017} and omit the treatment lead indicators associated with $e=0$ and with $e=-31$. Like \cite{grooms2015}, our specifications are weighted by total facilities in a state, and all standard errors are clustered at the state level.

Second, we make specific PTAs and use the new estimators described previously. Because our empirical application includes a set of ``never-treated'' states, we estimate event-study-type parameters based on the PTA (\ref{ass:CS-PTA1}) and use $\widehat{\delta}_{never}^{es}(e)$ as an estimator for $\delta^{es}(e)$. We also leverage the PTA (\ref{ass:CS-PTA2}) and use $\widehat{\delta}_{ny}^{es}(e)$ as an estimator for $\delta^{es}(e)$. Finally, we employ the PTA (\ref{ass:W-PTA}) and use $\widehat{\delta}_{ny+}^{es}(e)$ as an estimator for $\delta^{es}(e)$. We do not use the event-study estimates based on GMM framework discussed in Section \ref{sec:GMM}, since, in our specific application, the GMM associated with the PTA \ref{ass:dCdH-PTA} involves 780 moments with \emph{195 overidentification restrictions}, whereas sample size (state-year pairs) is equal to 759.

Next, we analyze whether the effect of state authorization on violation rates vary depending on whether a state has a long prevalence of corruption. To do so, we follow \cite{grooms2015} and consider the following TWFE specification
 \begin{equation}
Y_{it}=\alpha_{i}+\alpha_{t}+\sum_{e =-30, e\not=0}^{32}\beta _{e }1\left\{
t-G_{i}+1=e \right\} + \sum_{e =-30, e\not=0}^{32}\beta _{e}^{c} (1\left\{
t-G_{i}+1=e \right\}\times Corrupt_{i})+v_{it},
\label{eq:event2-app}
\end{equation}%
where the $\beta _{e}^{c}$'s are considered to be a measure of how treatment effects vary depending on whether a state is ``corrupt'' or not: positive (negative) point estimates suggest that the violation rates increased (decreased) more in corrupt states than in non-corrupt states.

At this stage, two important questions arise. First, what type of parallel trends assumption is actually being invoked to justify attaching a causal interpretation to the $\beta _{e}^{c}$'s in (\ref{eq:event2-app})? Second, is (\ref{eq:event2-app}) susceptible to the potential pitfalls discussed in Section \ref{sec:pit}? Answering these questions is inherently hard, as TWFE is a model specification and not a ``research design.'' An alternative, and perhaps more constructive way of approaching this problem is to construct event-study-type estimators that explicitly rely on a particular PTA, and that, by design, avoid the potential lack of a clear interpretation associated with the TWFE specification (\ref{eq:event2-app}). We follow this latter path. 

With respect to the PTA, there are two natural variants of each of the PTAs \ref{ass:CS-PTA1}, \ref{ass:CS-PTA2}, and \ref{ass:W-PTA} that one can invoke to highlight treatment effect heterogeneity with respect to whether a state is corrupt or not. These variants differ from each other depending on whether or not one allows for different counterfactual trends between corrupt and non-corrupt states. One may be concerned, for example, that, in the absence of treatment, the evolution of the violation rate could differ between corrupt and non-corrupt states. In this case, one would prefer a ``weaker'' assumption to allow for corruption-specific trends. In the context of our application, ``corruption'' is not randomly assigned and the geographic clustering of corrupt and non-corrupt states may lead us to prefer a PTA that permits corruption-specific trends if we think, for example, there may be regional trends that differ across these states. We believe this is the most natural identification set up, as this is how one would proceed if one were to separately estimate counterfactuals for corrupt states and non-corrupt states, and only later compare their difference. We formalize these six different PTAs below.

\begin{assumption}[Parallel trends assumption based on \textquotedblleft
never treated\textquotedblright\ units, with corruption-specific trends]
\label{ass:CS-PTA1-c1} For $c = {0,1}$, and all $g,t=2,\ldots ,\mathcal{T}$, such that $t\geq g$, 
\begin{equation*}
\mathbb{E}\left[ \left. Y_{t}\left( 0\right) -Y_{t-1}\left( 0\right)
\right\vert G_{g}=1, Corr = c\right] =\mathbb{E}\left[ \left. Y_{t}\left( 0\right)
-Y_{t-1}\left( 0\right) \right\vert C=1, Corr = c\right] .
\end{equation*}
\end{assumption}

\begin{assumption}[Parallel trends assumption based on \textquotedblleft
not-yet treated\textquotedblright\ units, with corruption-specific trends]
\label{ass:CS-PTA2-c1}For $c = {0,1}$, and all $g,s,t =2,\ldots ,\mathcal{T}$, such that $t\geq g$, $s\geq t$, 
\begin{equation*}
\mathbb{E}\left[ \left. Y_{t}\left( 0\right) -Y_{t-1}\left( 0\right)
\right\vert G_{g}=1, Corr = c\right] =\mathbb{E}\left[ \left. Y_{t}\left( 0\right)
-Y_{t-1}\left( 0\right) \right\vert D_{s}=0, Corr = c\right] .
\end{equation*}
\end{assumption}

\begin{assumption}[``Weaker'' Parallel trends assumption based on \textquotedblleft
not-yet treated\textquotedblright\ units, with corruption-specific trends]
\label{ass:W-PTA-c1} For $c = {0,1}$, and all $g,t =2,\ldots ,\mathcal{T}$, such that $t\geq g$, 
\begin{equation*}
\mathbb{E}\left[ \left. Y_{t}\left( 0\right) -Y_{t-1}\left( 0\right)
\right\vert G_{g}=1, Corr = c\right] =\mathbb{E}\left[ \left. Y_{t}\left( 0\right)
-Y_{t-1}\left( 0\right) \right\vert D_{t}=0, Corr = c\right] .
\end{equation*}
\end{assumption}

\begin{assumption}[Parallel trends assumption based on \textquotedblleft
never treated\textquotedblright\ units, without corruption-specific trends]
\label{ass:CS-PTA1-c2} For $c = {0,1}$, and all $g,t=2,\ldots ,\mathcal{T}$, such that $t\geq g$,
\begin{equation*}
\mathbb{E}\left[ \left. Y_{t}\left( 0\right) -Y_{t-1}\left( 0\right)
\right\vert G_{g}=1, Corr = c\right] =\mathbb{E}\left[ \left. Y_{t}\left( 0\right)
-Y_{t-1}\left( 0\right) \right\vert C=1\right] .
\end{equation*}
\end{assumption}

\begin{assumption}[Parallel trends assumption based on \textquotedblleft
not-yet treated\textquotedblright\ units, without corruption-specific trends]
\label{ass:CS-PTA2-c2}For $c = {0,1}$, and all $g,s,t =2,\ldots ,\mathcal{T}$, such that $t\geq g$, $s\geq t$,  
\begin{equation*}
\mathbb{E}\left[ \left. Y_{t}\left( 0\right) -Y_{t-1}\left( 0\right)
\right\vert G_{g}=1, Corr = c\right] =\mathbb{E}\left[ \left. Y_{t}\left( 0\right)
-Y_{t-1}\left( 0\right) \right\vert D_{s}=0\right] .
\end{equation*}
\end{assumption}

\begin{assumption}[``Weaker'' Parallel trends assumption based on \textquotedblleft
not-yet treated\textquotedblright\ units, without corruption-specific trends]
\label{ass:W-PTA-c2} For $c = {0,1}$, and all $g,t =2,\ldots ,\mathcal{T}$, such that $t\geq g$, 
\begin{equation*}
\mathbb{E}\left[ \left. Y_{t}\left( 0\right) -Y_{t-1}\left( 0\right)
\right\vert G_{g}=1, Corr = c\right] =\mathbb{E}\left[ \left. Y_{t}\left( 0\right)
-Y_{t-1}\left( 0\right) \right\vert D_{t}=0\right] .
\end{equation*}
\end{assumption}

The difference between these PTAs depends on whether one uses the ``never-treated'', some ``not-yet-treated'', or all ``not-yet-treated'' as valid comparisons group, and whether one only uses states with the same corruption status (corrupt or non-corrupt) as valid comparison groups. Assumptions \ref{ass:CS-PTA1-c1}-\ref{ass:W-PTA-c1}) do not assume that the evolution of the violation rate is the same between corrupt and non-corrupt states. These three assumptions are the analogues of Assumptions \ref{ass:CS-PTA1}, \ref{ass:CS-PTA2} and \ref{ass:W-PTA} when one restricts attention to the subset of units with corruption status equal to $c$. Assumptions \ref{ass:CS-PTA1-c2}-\ref{ass:W-PTA-c2}, on the other hand, assume that, in the absence of treatment, the evolution of the violation rate is the same for corrupt and non-corrupt states, i.e., it rules out corruption-specific trends. As so, one may argue that the Assumptions \ref{ass:CS-PTA1-c1}-\ref{ass:W-PTA-c1} are ``weaker'' than Assumptions \ref{ass:CS-PTA1-c2}-\ref{ass:W-PTA-c2}.

Next, one can easily leverage any of these PTAs to identify and estimate sensible treatment effect parameters by following the same steps described in Section \ref{sec:parameters}. The first step toward this goal is to show that the $ATT(g,t)$'s for the units with corruption status equal to $c$, $c=0,1$, defined by
\begin{equation*}
ATT\left( g,t;c\right) \equiv \mathbb{E}\left[ \left. Y_{it}\left( 1\right)
-Y_{it}\left( 0\right) \right\vert G_{g}=1, Corr = c\right],  \label{ATT_gt_c}
\end{equation*}
are nonparametrically point-identified for all $t\ge g$. However, given the results in Theorem 1 of C\&S and the discussion in Sections \ref{sec:parameters} and \ref{sec:Alternative}, this is a straightforward task. Indeed, one can easily show that for all $t\ge g$, the $ATT(g,t)$'s are nonparametrically identified by 
\begin{small}
\begin{eqnarray}
ATT_{never}^{1}\left( g,t;c\right)  &=&\mathbb{E}\left[ \left.
Y_{t}-Y_{g-1}\right\vert G_{g}=1,Corr = c\right] -\mathbb{E}\left[ \left.
Y_{t}-Y_{g-1}\right\vert C=1, Corr = c\right] ,\label{eq:att_never_c1} \\
ATT_{ny}^{1}\left( g,t;c\right)  &=&\mathbb{E}\left[ \left.
Y_{t}-Y_{g-1}\right\vert G_{g}=1, Corr = c\right] -\mathbb{E}\left[ \left.
Y_{t}-Y_{g-1}\right\vert D_{t}=0, Corr = c\right] \label{eq:att_ny_c1}\\
ATT_{ny+}^{1}\left( g,t;c\right)  &=&\mathbb{E}\left[ \left.
Y_{t}-Y_{g-1}\right\vert G_{g}=1, Corr = c\right] -\left(\sum_{s=g}^{t}\mathbb{E}\left[ \left.
\Delta Y_{s}\right\vert D_{s}=0, Corr = c \right] \right), \label{eq:att_ny+_c1}\\
ATT_{never}^{2}\left( g,t;c\right)  &=&\mathbb{E}\left[ \left.
Y_{t}-Y_{g-1}\right\vert G_{g}=1,Corr = c\right] -\mathbb{E}\left[ \left.
Y_{t}-Y_{g-1}\right\vert C=1\right] , \label{eq:att_never_c2}\\
ATT_{ny}^{2}\left( g,t;c\right)  &=&\mathbb{E}\left[ \left.
Y_{t}-Y_{g-1}\right\vert G_{g}=1, Corr = c\right] -\mathbb{E}\left[ \left.
Y_{t}-Y_{g-1}\right\vert D_{t}=0\right] \label{eq:att_ny_c2},\\
ATT_{ny+}^{2}\left( g,t;c\right)  &=&\mathbb{E}\left[ \left.
Y_{t}-Y_{g-1}\right\vert G_{g}=1, Corr = c\right] -\left(\sum_{s=g}^{t}\mathbb{E}\left[ \left.
\Delta Y_{s}\right\vert D_{s}=0\right] \right) \label{eq:att_ny+_c2},
\end{eqnarray}
\end{small}depending on whether one respectively imposes one of the PTA \ref{ass:CS-PTA1-c1}-\ref{ass:W-PTA-c2}.\footnote{Although the PTAs \ref{ass:CS-PTA1-c2}-\ref{ass:W-PTA-c2} lead to overidentification, for the sake of simplicity we do not fully exploit all these restrictions  when proposing the aforementioned estimands.} These results are analogous to (\ref{eq:att_never_pop}), (\ref{eq:att_ny_pop}) and (\ref{eq:att_ny_weak}). Likewise, all the aforementioned quantities can be estimated using the analogy principle, i.e., by replacing population expectation with sample expectations.

Armed with these estimators, one can form different summary measures for the overall treatment effect following the same steps described in Section \ref{sec:parameters}. To explicitly show how one can form event-study-type estimators, let ${ATT}_{generic}\left( g,t;c\right)$ be a generic notation for ${ATT}_{never}^{1}\left( g,t;c\right)$, ${ATT}_{ny}^{1}\left( g,t;c\right)$, ${ATT}_{ny+}^{1}\left( g,t;c\right)$, ${ATT}_{never}^{2}\left( g,t;c\right)$, ${ATT}_{ny}^{2}\left( g,t;c\right)$, and ${ATT}_{ny+}^{2}\left( g,t;c\right)$ and denote its plug-in estimator by $\widehat{ATT}_{generic}\left( g,t;c\right)$. Then, one can estimate the average treatment effect for units with corruption status equal to $c$ that have
been treated for $e$ periods by 
\begin{equation}
\widehat{\delta}^{es}_{generic}\left( e; c\right) =\sum_{g=2}^{\mathcal{T}}\sum_{t=2}^{\mathcal{T}%
}1\left\{ t-g+1=e\right\} \widehat{w}\left(g;e,c\right) \widehat{{ATT}}_{generic}\left( g,t;c\right) ,  \label{eq:event-study-est-c}
\end{equation}%
where the weights are given by
\begin{equation*}
\widehat{w}\left(g;e,c\right) \equiv \widehat{P}\left( G_{g}=1|\text{Treated for $\ge$ e periods, Corr = c}\right) = \frac{N_{g\cap \geq e \cap c}}{%
N_{\geq e \cap c}},
\label{eq:dyn-weights-c}
\end{equation*}%
$N_{g\cap \geq e \cap c}$ denotes the number of observations in group $g$ among
those units with corrupt status $c$ that have been treated for at least $e$ periods, and $N_{\geq e}$ is
the number of units with corrupt status $c$ who have been treated for at least $e$ periods.\footnote{When $e<0$, we replace $\widehat{w}\left(g;e,c\right)$ in (\ref{eq:event-study-est-c}) with $\widehat{w}\left(g;e-,c\right)$, where $\widehat{w}\left(g;e-,c\right)$ is analogous to $\widehat{w}\left(g;e-\right)$ as defined in Remark \ref{rm:pre-test}.} Given that our main goal is to compare the evolution of treatment effects between corrupt and non-corrupt states, we can simply compute the difference between  $\widehat{\delta}^{es}_{generic}\left( e; 1\right)$ and $\widehat{\delta}^{es}_{generic}\left( e; 0\right)$. Denote this (generic) estimator by $\widehat{\delta}^{es}_{generic}\left( e; 1-0\right)$, i.e., 
\begin{equation}
\widehat{\delta}^{es}_{generic}\left( e; 1-0\right) = \widehat{\delta}^{es}_{generic}\left( e; 1\right) - \widehat{\delta}^{es}_{generic}\left( e; 0\right). \label{eq:event-study-est-diff}
\end{equation}

Here, we stress that, regardless of which of the six different estimators for (\ref{eq:event-study-est-diff}) one adopts, they are all directly and explicitly tied to a given PTA, and, by design, they bypass the potential pitfalls associated with the TWFE specification.

In addition to the event-study estimates, we further aggregate these treatment effect curves into scalar, easy to interpret parameters. Toward this end, we report the plug-in estimators for the following two aggregated treatment effect parameters proposed by C\&S,
\begin{equation}
ATT^{simple, 1-0} = ATT^{simple;1} - ATT^{simple;0} \label{eqn:simplesum_het},%
\end{equation}
where, for $c=0,1$ $ATT^{simple;c}$ is defined analogously to (\ref{eqn:simplesum}), i.e.,
\begin{equation*}
ATT^{simple, c} = \frac
{\sum_{g=2}^{\mathcal{T}}\sum_{t=2}^{\mathcal{T}}\mathbf{1}\{g\leq
t\}P(G=g, Corr = c) \cdot ATT(g,t;c)}{\sum_{g=2}^{\mathcal{T}}\sum_{t=2}^{\mathcal{T}}\mathbf{1}\{g\leq
t\}P(G=g, Corr = c)} \label{eqn:simplesum_c},%
\end{equation*}
and the average of $\delta ^{es}\left( e; 1-0\right)$ over all possible (positive) values of $e$,
\begin{equation}
\delta^{e, avg, 1-0}=\delta^{e, avg; 1} - \delta^{e, avg; 0}. \label{eqn:dyn_c}%
\end{equation}
where, for $c=0,1$, $\delta^{e, avg;c}$ is defined analogously to (\ref{eqn:dyn}), i.e.,
\begin{equation*}
\delta^{e, avg;c}=\frac{1}{\mathcal{T}-1}\sum_{e=1}^{\mathcal{T}-1}\delta^{es}%
(e;c). 
\end{equation*}
We report estimators for these functionals that rely on the PTA \ref{ass:CS-PTA1-c1}-\ref{ass:W-PTA-c2}, respectively. For the sake of comparison, we also report the OLS estimate of $\beta_{fe}^{c}$ associated with the following TWFE specification,
 \begin{equation}
 Y_{it}=\alpha _{g}+\alpha _{t}+\beta _{fe}D_{it}+\beta _{fe}^{c}D_{it}\times Corrupt_{i}+ u_{it},
\label{eq:TWFE-app-int}
\end{equation}%
though these estimates are also subject to pitfalls briefly described in Section \ref{sec:pit}.
\subsection{Results}
\subsubsection{Baseline results}

\begin{figure}[htp]
\centering
\caption{Event-study analysis of violation rate: baseline results} \label{fig:event1}
\includegraphics[width=.9\linewidth]{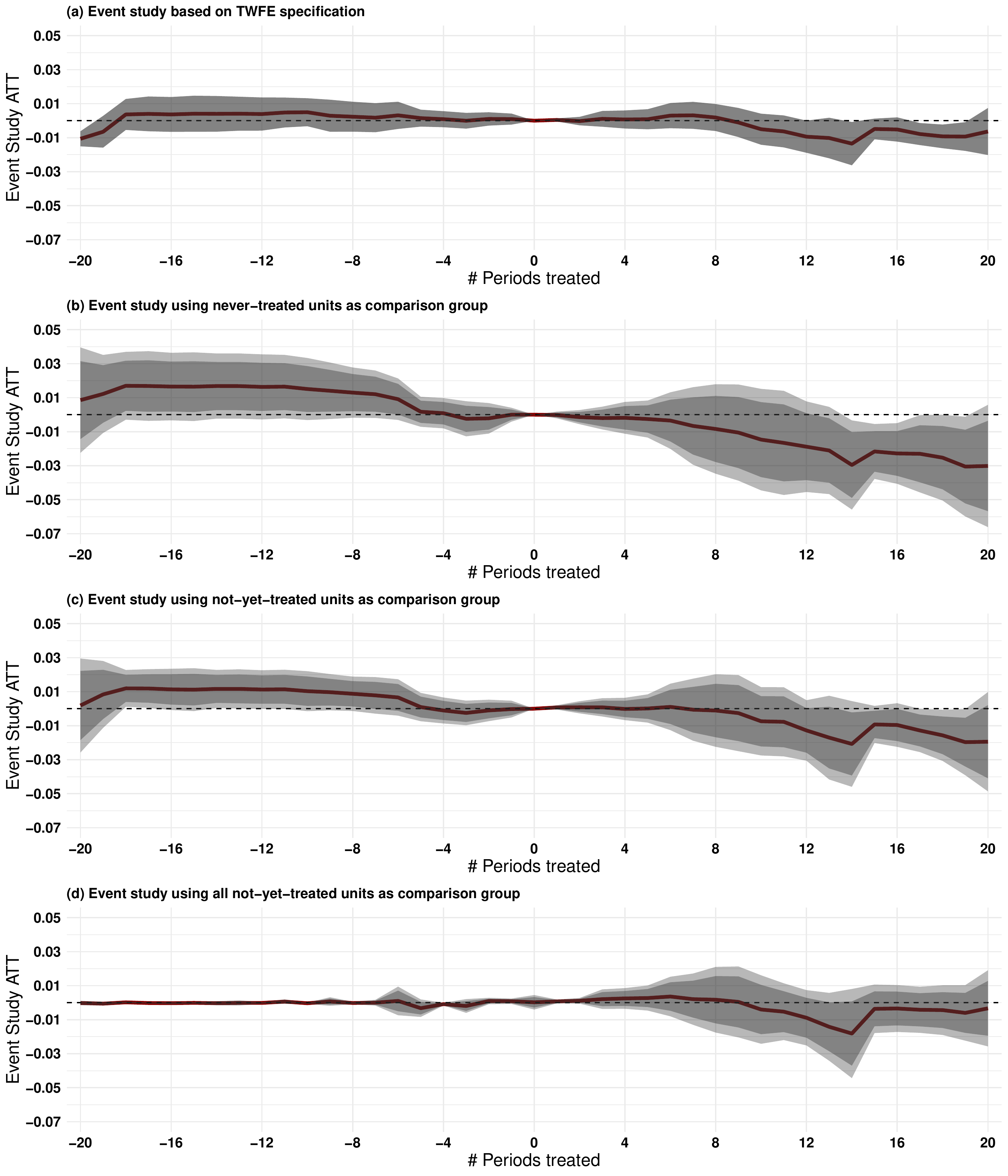} \\
\caption*{\footnotesize Notes: Red line displays the point estimate, dark-shaded area the 90\% pointwise confidence interval, and the light-shaded are the 90\% simultaneous confidence band. Panel (a) displays the ordinary least squares (OLS) estimates of the $\beta_{e}$ associated with the two-way fixed-effects linear regression specification (\ref{eq:event1-app}); Panel (b) displays the results based on (\ref{eq:event-study}) that uses (\ref{eq:att_never}) as an estimator for $ATT(g,t)$; Panel (c) displays the results based on (\ref{eq:event-study}) that uses (\ref{eq:att_ny}) as an estimator for $ATT(g,t)$; Panel (d) displays the results based on (\ref{eq:event-study}) that uses (\ref{eq:att_ny+}) as an estimator for $ATT(g,t)$. All standard errors are clustered at the state level, though the standard errors in Panel (a) are based on analytical results, whereas those in Panel (b)-(d) are based on the multiplicative bootstrap procedure discussed in Algorithm~B.1) and in C\&S (we use 1,000 bootstrap draws). The critical value for the simultaneous confidence bands is computed using Algorithm~B.1 (which is akin to the one proposed by C\&S).}
\end{figure}

Figure~\ref{fig:event1} displays the results based on the TWFE specification (\ref{eq:event1-app}), and those based on the event-study estimators $\widehat{\delta}_{never}^{es}(e)$, $\widehat{\delta}_{ny}^{es}(e)$, and $\widehat{\delta}_{ny+}^{es}(e)$. We report the point-estimates associated with 20 treatment leads and 20 treatment lags (red line), their associated 90\% point-wise confidence intervals (dark-shaded area), and 90\% \emph{simultaneous} confidence intervals (light-shaded area) --- we do not report simultaneous confidence intervals for the TWFE specification as these are usually not reported by practitioners who adopt such specifications. It is important to emphasize that, in each of the panels in Figure~\ref{fig:event1}, we have 40 different estimates, one for each considered $e$. Point-wise inference procedures proceed ``as if'' one were conducting a single hypothesis test, and report standard confidence interval for each $e$. Failing to account for the fact that one is performing 40 \emph{different} hypotheses tests may lead to significant treatment effects and/or pre-trends that emerge simply by chance. Simultaneous confidence intervals, on the other hand, account for this multiple testing problem, and asymptotically cover the entire event-study curve with probability 1 - $\alpha$, where $\alpha$ is the significance level. As so, simultaneous confidence intervals are suitable to analyze global properties of the event-study curve, such as monotonicity and presence of statistically non-zero effects. In practice, one simply has to replace the commonly used critical value (say, 1.645 for a 90\% confidence interval) with the one simulated via a bootstrap procedure akin to Algorithm~B.1; see Section 4 of C\&S for additional details.

The results shown in Figure~\ref{fig:event1} suggest that, regardless of the PTA and the estimator used, there is little to no evidence that the transition to state control decreased violation rates. Despite the similarity in terms of conclusions, we find that comparing the results from each specification highlights some interesting practical features. For instance, the point estimates associated with the TWFE specification (Panel (a)) and with the estimator that uses the ``all not-yet-treated'' states as a comparison group (Panel (d)) are close to each other, whereas using 
``never-treated'' states as a comparison group (Panel (b)) suggests a slightly stronger long-run effect. Furthermore, when using  ``all not-yet-treated'' states as a comparison group (Panel (d)), the (simultaneous) confidence interval is tighter, suggesting, as we discussed in Section \ref{sec:Alternative}, that it makes more efficient use of the available data. In terms of interpreting the pre-treatment coefficients, the pre-trend point estimates when using the ``not-yet-treated'' comparison group in Panel (c) are closer to zero than when one uses the ``never-treated'' states as a comparison group in Panel (b). It is also very noticeable that pre-treatment trends in Panel (d) are very precisely estimated zeros. However, it is important to recall from Remark \ref{rem:new-pre-trends-est} that these pre-treatment coefficients should not be directly compared to the other pre-treatment trends as they measure ``local deviations'' of zero pre-treatment trends rather ``cumulative deviations'' of pre-treatment trends.

Although these estimators lead to similar conclusions, they are not (a priori) ``made equal''. As highlighted by S\&A and discussed in Section \ref{sec:pit}, the $\beta_e$'s associated with the TWFE specification (\ref{eq:event1-app}) are not guaranteed to have a clear causal interpretation, \emph{even when one invokes the PTA \ref{ass:dCdH-PTA}}, which, in our application, imposes 195 overidentifying restrictions on the evolution of violation rates across states. The estimators in Panels (b), (c), and (d), on the other hand, are designed to bypass the potential pitfalls of the TWFE specification, and rely on clearly stated parallel trends assumptions (Assumptions \ref{ass:CS-PTA1}, \ref{ass:CS-PTA2}, and \ref{ass:W-PTA}, respectively).

\begin{table}[htb]
	\begin{center}
	\caption{Effect of authorization on violation rate: Baseline specifications.}	
	\label{tab:baseline}
	\begin{adjustbox}{ max width=1\linewidth, max totalheight=1\textheight, keepaspectratio}
\begin{threeparttable}

\begin{tabular}{@{}lccccccc@{}} \noalign{\vskip 2mm} 
\hline \hline

\noalign{\vskip 2mm} Summary measures \phantom{abcdefgh} & \phantom{ab}Never-treated\phantom{ab} & \phantom{ab} & \phantom{ab}Not-yet-treated\phantom{ab} & \phantom{ab} & \phantom{ab}All Not-yet-treated\phantom{ab} \phantom{ab} & \phantom{ab} & TWFE\phantom{ab} \\ 
& (1) &  & (2) & & (3) & & (4) \\ \hline 

\noalign{\vskip 2mm} $ATT^{simple}$ & -0.017 & & -0.010&  & -0.003& & --- \\
                               & (0.009) & & (0.009)& & (0.006)& &--- \\
                               & [-0.032, 0.001] & & [-0.024, 0.004]& & [-0.014, 0.008]& &--- \\
[1em]
$\delta^{e, avg}$ & -0.015 & & -0.008 & & -0.003 & &--- \\
                                         & (0.007) & & (0.006) & & (0.004) & &--- \\
                                         & [-0.027, -0.002] & & [-0.017, 0.002]& & [-0.010, 0.004]& &---  \\
[1em]                                         
TWFE & --- & & --- & & --- & &-0.003\\
     & --- & &--- & & --- & & (0.010)\\
     & --- & &---&  & --- & & [-0.019, 0.013]\\
\bottomrule
\end{tabular}%

\begin{tablenotes}[para,flushleft]
\small{
Notes: The point estimates, cluster-robust standard errors (in parenthesis), and 90\% confidence interval (in brackets) for the effect of state authorization on violation rates. $ATT^{simple}$ is as defined in (\ref{eqn:simplesum}) and denotes the weighted average of all post-treatment $ATT(g,t)'s$. $\delta^{e, avg}$ is as defined in (\ref{eqn:dyn}) and denotes the time-average of all event-study parameters $\delta^{es}(e)$, $e>0$. TWFE refers to the ordinary least square estimates of $\beta_{fe}$ in the TWFE linear regression specification (\ref{eq:TWFE-1}), which is invariant to the comparison group being used. Column (1) display the results that uses (\ref{eq:att_never}) as an estimator for $ATT(g,t)$,  column (2)  displays the results that uses (\ref{eq:att_ny}) as an estimator for $ATT(g,t)$, and   column (3)  displays the results that uses (\ref{eq:att_ny+}) as an estimator for $ATT(g,t)$ Column (4) displays the result using the TWFE regression specification.
Standard errors are clustered at the state level, and, with the exception of the TWFE summary measure, are computed using the multiplicative bootstrap procedure described in Algorithm~B.1, which is akin to the one proposed by C\&S. We use 1,000 bootstrap draws.
}
\end{tablenotes}
\end{threeparttable}
\end{adjustbox}

	\end{center}
\end{table}	

As discussed in Section \ref{sec:parameters}, there are multiple sensible measures that one can use to summarize the overall effect of state authorization on violation rates across all treated states. For instance, one can use the ``simple'' average of all the $ATT(g,t)$ where $t\ge g$, $ATT^{simple}$, as defined in (\ref{eqn:simplesum}), or the average of the event-study-type estimands $\delta ^{es}\left( e\right)$ over the positive values of $e$, $\delta^{e, avg}$, as defined in \ref{eqn:dyn}. Table~\ref{tab:baseline} shows the estimates of these parameters when one adopts Assumption \ref{ass:CS-PTA1} (Column (1)),  Assumption \ref{ass:CS-PTA2} (Column (2)), or Assumption \ref{ass:W-PTA} (Column (3)). For the sake of comparison we also report the OLS estimate of $\beta_{fe}$ (Column (4)). Standard errors, clustered at the state level, are reported in parenthesis and 90\% confidence intervals are reported in brackets. Essentially, all these summary measures indicates that state authorization has close to zero effect on violation rates, which is in line with the findings from \cite{grooms2015}.

\subsubsection{Corrupt vs. Non-corrupt Results}
\begin{figure}[htp]
\centering
\caption{Event-study analysis of violation rate: difference between corrupt and non-corrupt states allowing different counterfactual trends between corrupt and non-corrupt states} \label{fig:event2}
\includegraphics[width=.9\linewidth]{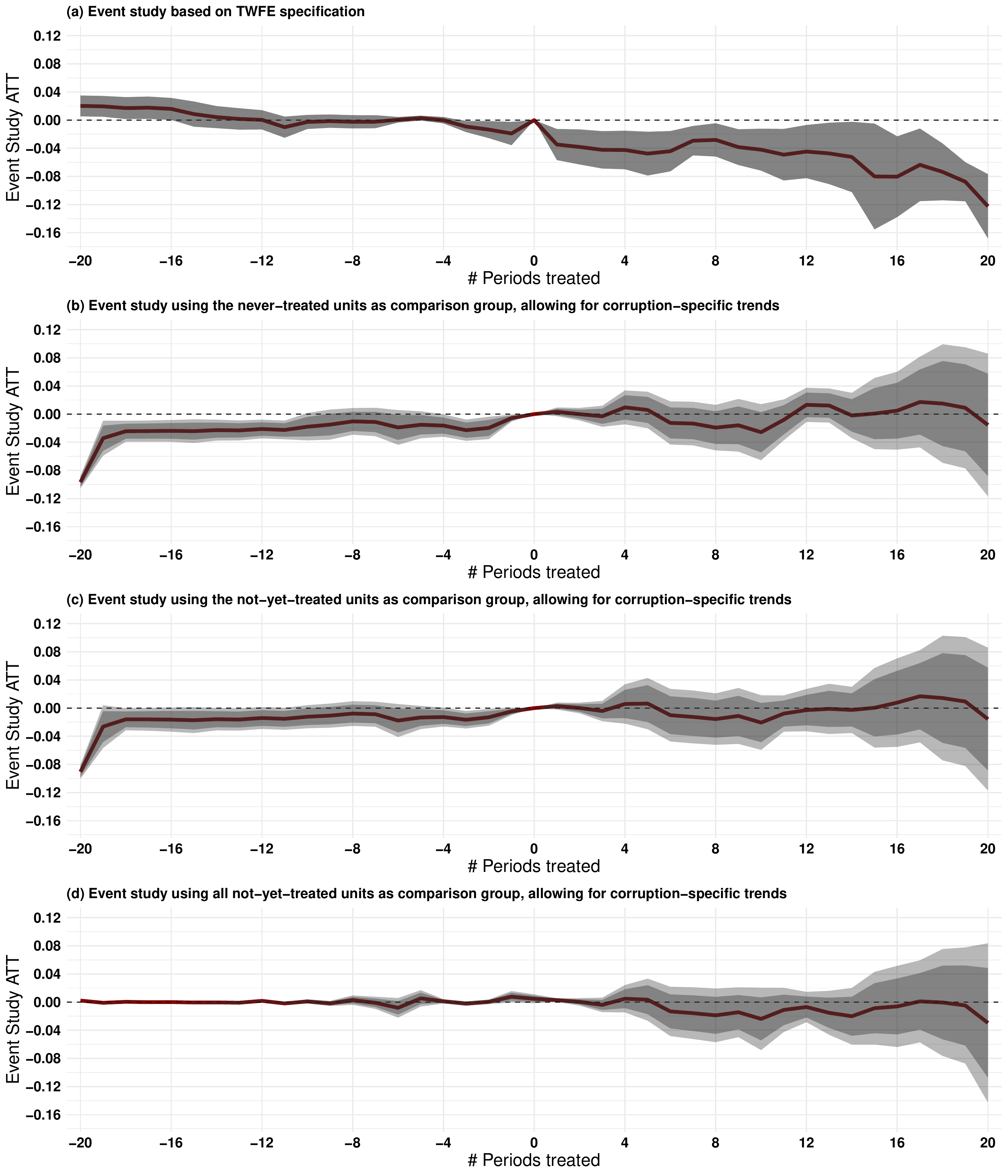} \\
\caption*{\footnotesize Notes: Red line displays the point estimate, dark-shaded area the 90\% pointwise confidence interval, and the light-shaded are the 90\% simultaneous confidence band. Panel (a) displays the results based on the OLS estimates of the $\beta _{e}^{c}$'s in the TWFE specification (\ref{eq:event2-app}); Panel (b) displays the results based on the event-study estimator (\ref{eq:event-study-est-c}) that relies on the PTA \ref{ass:CS-PTA1-c1}; Panel (c) displays the results based on the event-study estimator (\ref{eq:event-study-est-c}) that relies on the PTA \ref{ass:CS-PTA2-c1}; Panel (d) displays the results based on the event-study estimator (\ref{eq:event-study-est-c}) that relies on the PTA \ref{ass:W-PTA-c1}. All standard errors are clustered at the state level, though the standard errors in Panel (a) are based on analytical results, whereas those in Panel (b)-(d) are based on the multiplicative bootstrap procedure discussed in Algorithm~B.1, which is similar to C\&S proposal (we use 1,000 bootstrap draws). The critical value for the simultaneous confidence bands is computed using Algorithm~B.1.}
\end{figure}

Next, we analyze whether the effect of state authorization on violation rates vary depending on whether a state has a long prevalence of corruption. Panel (a) of Figure~\ref{fig:event2} displays the OLS estimates of   $\beta _{e}^{c}$'s, together with the 90\% pointwise confidence intervals. All standard errors are clustered at the state level. Consistent with the findings from \cite{grooms2015}, the results suggest that states with high levels of corruption have a lower violation rate after authorization relative to non-corrupt states, and the relative drop in the violation rate appears to increase with elapsed treatment time.

\begin{figure}[htp]
\centering
\caption{Event-study analysis of violation rate: difference between corrupt and non-corrupt states assuming same counterfactual trends for corrupt and non-corrupt states} \label{fig:event3}
\includegraphics[width=.9\linewidth]{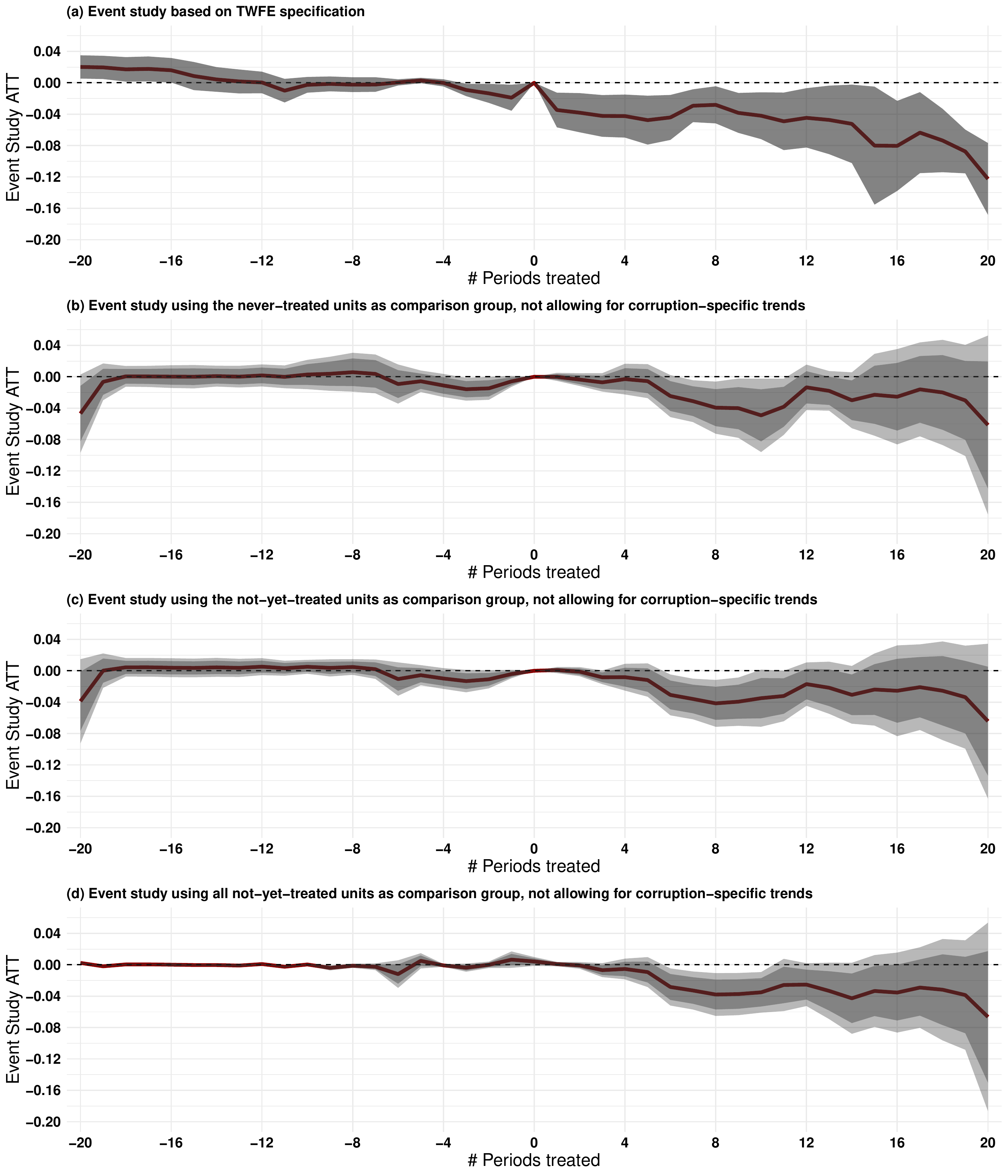} \\
\caption*{\footnotesize Notes: Red line displays the point estimate, dark-shaded area the 90\% pointwise confidence interval, and the light-shaded are the 90\% simultaneous confidence band. Panel (a) displays the results based on the OLS estimates of the $\beta _{e}^{c}$'s in the TWFE specification (\ref{eq:event2-app}); Panel (b) displays the results based on the event-study estimator (\ref{eq:event-study-est-c}) that relies on the PTA \ref{ass:CS-PTA1-c2}; Panel (c) displays the results based on the event-study estimator (\ref{eq:event-study-est-c}) that relies on the PTA \ref{ass:CS-PTA2-c2}; Panel (d) displays the results based on the event-study estimator (\ref{eq:event-study-est-c}) that relies on the PTA \ref{ass:W-PTA-c2}. All standard errors are clustered at the state level, though the standard errors in Panel (a) are based on analytical results, whereas those in Panel (b)-(d) are based on the multiplicative bootstrap procedure discussed in Algorithm~B.1, which is similar to C\&S proposal (we use 1,000 bootstrap draws). The critical value for the simultaneous confidence bands is computed using Algorithm~B.1.}
\end{figure}

 Panels (b), (c), and (d) of Figure~\ref{fig:event2} present the event-study estimates (\ref{eq:event-study-est-diff}) based on the PTAs \ref{ass:CS-PTA1-c1}, \ref{ass:CS-PTA2-c1}, and \ref{ass:W-PTA-c1} that allow for corruption-specific trends, whereas Panels (b), (c), and (d) of Figure~\ref{fig:event3} present the event-study estimates (\ref{eq:event-study-est-diff}) based on the PTAs \ref{ass:CS-PTA1-c2}, \ref{ass:CS-PTA2-c2}, and \ref{ass:W-PTA-c2} that do not allow for corruption-specific trends. For comparison purposes, Panel (a) of Figures~\ref{fig:event2} and~\ref{fig:event3} displays the OLS estimates of the  $\beta _{e}^{c}$'s associated with the TWFE specification (\ref{eq:event2-app}). Like before, all estimators are weighted by total facilities in a state, all standard errors are clustered at the state level, and we report both pointwise and simultaneous 90\% confidence intervals.

The results in Figures~\ref{fig:event2} and \ref{fig:event3} reveal the practical relevance of being explicit about the underlying PTA in a given application. For instance, Figure~\ref{fig:event2} suggests that when one invokes Assumption \ref{ass:CS-PTA1-c1} (Panel (b)), Assumption \ref{ass:CS-PTA2-c1} (Panel (c)), or Assumption \ref{ass:W-PTA-c1} (Panel (d)) and allows corruption-specific counterfactual trends, one finds that state authorization affected violation rates in a similar manner, regardless of whether a state is corrupt or not. Such results are in sharp contrast with those associated with the TWFE specification (\ref{eq:event2-app}) in Panel (a). 

On the other hand, the results in Figure~\ref{fig:event3} suggest that when one invokes Assumption \ref{ass:CS-PTA1-c2} (Panel (b)), Assumption \ref{ass:CS-PTA2-c2} (Panel (c)), or Assumption \ref{ass:W-PTA-c2} (Panel (d)) and therefore rules out differential potential trends between corrupt and non-corrupt states, one finds that corrupt states experience a large decrease in violation rates after state authorization relative to non-corrupt states. Although such results are in line with those based on TWFE in Panel (a), we note that the results based on (\ref{eq:event-study-est-diff}) suggests milder treatment effect differences between corrupt and non-corrupt states.

Next, we aggregate these treatment effect curves into a scalar, easy to interpret parameter. Table~\ref{tab:corrupt} reports the plug-in estimators for the two aggregated treatment effect parameters, shown in (\ref{eqn:simplesum_het}) and (\ref{eqn:dyn_c}). For comparison, we report in Column (7) the OLS estimate of $\beta_{fe}^{c}$ associated with the TWFE specification, shown in (\ref{eq:TWFE-app-int}).

\begin{table}[htp]

	\begin{center}
	\caption{Effect of authorization on violation rate: corrupt vs. not corrupt states.}	
	\label{tab:corrupt}
\begin{adjustbox}{ max width=1\linewidth, max totalheight=1\textheight, keepaspectratio}
\begin{threeparttable}

\begin{tabular}{@{}lccccccccc@{}} \hline \hline
   \noalign{\vskip 2mm}\phantom{a} & \multicolumn{3}{c}{Allow for corrupt-specific trends} & \phantom{ab} 
  & \multicolumn{3}{c}{Not allow corrupt-specific trend} & & \\ \cline{2-4} \cline{6-8} 
  
   \noalign{\vskip 2mm} Summary measures \phantom{abc} & \multicolumn{1}{l}{Never-treated} & \multicolumn{1}{l}{Not-yet-treated} &
   \multicolumn{1}{l}{All Not-yet-treated} & & \multicolumn{1}{l}{Never-treated} & \multicolumn{1}{l}{Not-yet-treated} &
   \multicolumn{1}{l}{All Not-yet-treated} & &TWFE \\ 
   & (1) & (2) & (3) & & (4) & (5) & (6) & & (7)\\
   \hline 
  
 \noalign{\vskip 2mm} $ATT^{simple, 1-0}$   & -0.007 & -0.008 & -0.014 &  & -0.035 & -0.035 & -0.033 & &---\\
            & (0.014) & (0.014) & (0.013) &   & (0.012) & (0.012) & (0.010) & &---\\
            & [-0.030, 0.017] & [-0.031, 0.016] & [-0.036, 0.008] &   & [-0.054, -0.016] & [-0.054, -0.015] & [-0.049, -0.017] & &---\\
[1em]
$\delta^{e, avg,1-0}$   & -0.001 & -0.002 & -0.009 &   & -0.024 & -0.025 & -0.028 & &---\\
                        & (0.014) & (0.016) & (0.014) &   & (0.013) & (0.012) &(0.011) & &---\\
                        & [-0.024, 0.022] & [-0.028, 0.024] &  [-0.031, 0.013] &  & [-0.045, -0.003] & [-0.045, -0.005] & [-0.047, -0.009] & &--- \\
[1em]
TWFE    & --- & --- & --- &  & --- & --- & --- & & -0.037\\
        & --- & --- &  --- & & --- & --- & --- & &(0.010)\\
        & --- & --- & --- &  & --- & --- & --- & &[-0.054, -0.020]\\
\bottomrule
\end{tabular}%

\begin{tablenotes}[para,flushleft]
\small{
Notes: The point estimates, cluster-robust standard errors (in parenthesis), and 90\% confidence interval (in brackets) for the effect of state authorization on violation rates. $ATT^{simple, 2-0}$ is as defined in (\ref{eqn:simplesum_c}) and denotes the difference of the weighted average of all post-treatment $ATT(g,t;c)'s$ between corrupt and non-corrupt states.  $\delta^{e, avg, 1-0}$ is as defined in (\ref{eqn:dyn_c}) and denotes difference of the time-average of all event-study parameters $\delta^{es}(e)$, $e>0$, between corrupt and non-corrupt states. TWFE refers to the ordinary least square estimates of $\beta_{fe}^c{}$ in the TWFE linear regression specification (\ref{eq:TWFE-app-int}), which is invariant to the comparison group being used. Columns (1)-(6) display the results that relies on the PTA \ref{ass:CS-PTA1-c1}-\ref{ass:W-PTA-c2}, respectively. Standard errors are clustered at the state level, and, with the exception of the TWFE summary measure, are computed using the multiplicative bootstrap procedure presented in Algorithm B.1, which is akin to C\&S proposal. We use 1,000 bootstrap draws.

}
\end{tablenotes}
\end{threeparttable}
\end{adjustbox}

	\end{center}
\end{table}	
The results in Table~\ref{tab:corrupt} reinforce the message from  Figures~\ref{fig:event2} and \ref{fig:event3}: when one allows for corruption-specific trends and relies on PTAs \ref{ass:CS-PTA1-c1}, \ref{ass:CS-PTA2-c1}, or \ref{ass:W-PTA-c1} (Columns (1), (2), and (3), respectively), one finds essentially no evidence that the effect of state authorization on violation rates varies by state corruption. On the other hand, when one relies on the ``stronger'' PTAs \ref{ass:CS-PTA1-c2}, \ref{ass:CS-PTA2-c2}, or \ref{ass:W-PTA-c2} (Columns (4), (5), and (6), respectively), one finds evidence that corrupt states experienced a large decrease in violation rate after state authorization relative to non-corrupt states. This latter result is in agreement with the TWFE specification, whereas the former is not.

\section{Conclusion\label{sec:Conclusion}}
In this paper, we have highlighted the important role played by the parallel trends assumption in event-study settings in terms of identification, estimation and summary of different treatment effects parameters. We first showed that, when there is variation in treatment timing, researchers may adopt different types of parallel trends assumptions and identify/estimate different treatment effect parameters. Next, we discussed the practical implications of adopting different parallel trends assumptions, and discussed how one constructs estimators that make use of all the restrictions implied by the underlying PTA. Here, we documented an interesting ``robustness" vs. ``efficiency" trade-off in terms of the strength of the underlying PTA, and argue that one should take this into consideration whenever employing a DID-type of analysis. Importantly, we advocate that one should always attempt to be explicit about the parallel trends assumption invoked in the study, as this usually translates into a more transparent and objective analysis. We showed how one can form semiparametrically efficient DID estimators by fully exploiting all the empirical content of underlying PTA via the traditional GMM approach. We also proposed an alternative, simpler to use DID estimator that does not restrict pre-treatment trends when one wants to use ``not-yet-treated'' units as a comparison group, and, at the same time, makes use of more groups than other available DID estimators. Finally, we illustrated the practical importance of being explicit about the PTA via an empirical application about the effect of the transition from federal to state management of the Clean Water Act on compliance rates. Our results suggest that the conclusion that corrupt states see a decline in the violation rate after program authorization relative to non-corrupt treated states depends on the type of PTA adopted. 

\singlespacing
\bibliography{bibliography}


\end{document}